\documentclass[11pt,aps,floatfix]{revtex4}
\usepackage{amssymb}
\usepackage{latexsym}

\usepackage{epsfig}
\usepackage{psfrag}

\usepackage{dcolumn}
\usepackage{amsmath}
\usepackage{amsfonts}
\usepackage{bm}
\usepackage{color}

\makeatletter

\@ifundefined{textcolor}{}
{%
 \definecolor{BLACK}{gray}{0}
 \definecolor{WHITE}{gray}{1}
 \definecolor{RED}{rgb}{1,0,0}
 \definecolor{GREEN}{rgb}{0,1,0}
 \definecolor{BLUE}{rgb}{0,0,1}
 \definecolor{CYAN}{cmyk}{1,0,0,0}
 \definecolor{MAGENTA}{cmyk}{0,1,0,0}
 \definecolor{YELLOW}{cmyk}{0,0,1,0}
}
\makeatother

\newcommand{\be}{\begin{equation}}
\newcommand{\ee}{\end{equation}}
\newcommand{\bea}{\begin{eqnarray}}
\newcommand{\eea}{\end{eqnarray}}

\newcommand{\p}{\partial}
\newcommand{\s}{\sigma}

\newcommand{\la}{\langle}
\newcommand{\ra}{\rangle}

\newcommand{\ri}{\mbox{i}}
\newcommand{\re}{\mbox{e}}

\begin{document}

\title{Studying the Perturbed Wess-Zumino-Novikov-Witten $SU(2)_k$ Theory Using the Truncated Conformal Spectrum Approach}
\author{R. M. Konik$^1$, T. P\'almai$^2$, G. Tak\'acs$^{2,3}$, and A. M. Tsvelik$^1$}
\affiliation{ $^1$Department of Condensed Matter Physics and Materials Science, \\
Brookhaven National Laboratory,
  Upton, NY 11973-5000, USA;\\
$^2$MTA-BME ''Momentum´´ Statistical Field Theory Research Group,\\
H-1111 Budapest, Hungary, Budafoki \'ut 8.
\\
$^3$Department of Theoretical Physics,
Institute of Physics, \\
Budapest Univ. Technology and Economics,
H-1111 Budapest, Hungary, Budafoki \'ut 8.} 
\date{\today } 
\begin{abstract} 
We study the  $SU(2)_k$ Wess-Zumino-Novikov-Witten (WZNW) theory perturbed 
by the trace of the primary field in the adjoint representation, a theory governing the low-energy 
behaviour of a class of strongly correlated electronic systems. While
the model is non-integrable, its dynamics can be investigated using the numerical technique 
of the truncated conformal spectrum approach combined with numerical and analytical renormalization groups (TCSA+RG). The numerical results so obtained provide
support for a semiclassical analysis valid at $k\gg 1$. Namely, we find that the low energy behavior is sensitive to the sign of the coupling constant, $\lambda$.  
Moreover for $\lambda >0$ 
this behavior depends on whether $k$ is even or odd.  With $k$ even, we find definitive evidence that the model at low energies is equivalent to the massive O(3) sigma model.
For $k$ odd, the numerical evidence is more equivocal, but we find indications that the low energy effective theory is critical.

\end{abstract}


\maketitle
\section{Introduction}

Conformal field theories (CFT) describe universal critical behavior and by virtue of this play an enormously important role in 
the physics of strongly correlated systems. This universality is not completely lost in the presence of perturbations since, as a rule,  
the number of relevant operators is finite and once restricted by symmetry, often number but a few.  Physics of perturbed critical models can be 
rich and complex, especially when the perturbation is non-integrable.  For examples one may look at the quantum Ising model 
perturbed simultaneously by a longitudinal magnetic field and the thermal operator \cite{fonseca,ising_md} or double sine-Gordon models \cite{double1, double2}. 

A focus on relevant perturbations of a CFT is most appropriate when the perturbations are strongly relevant. 
Indeed, the more relevant the perturbation the smaller is the energy scale over which the spectrum is significantly altered. This feature lies 
at the foundation of the truncated conformal spectrum approach (TCSA) introduced in \cite{truncated}. In the simplest version of this 
approach (TCSA), one truncates the spectrum of the unperturbed CFT which reduces the problem to numerical diagonalization of finite 
size matrices. Later this idea was combined with a numerical renormalization group \cite{tsrg} (TCSA+NRG).  The TCSA+NRG
has been used to tackle a number of problems ranging from the excitonic spectrum in semiconducting carbon nanotubes 
\cite{konik_scnt1,konik_scnt2}, to
quenches in the Lieb-Liniger model \cite{jsc_rmk,bran_jsc_rmk}, to studying theories whose fields live on a non-compact manifold \cite{lg}.
In a further development, the precision of TCSA or TCSA+NRG computations can be improved further upon using perturbative renormalization 
group techniques \cite{watts_arxiv,watts,konik_scnt1,takacs,rychkov,konik_scnt2}.   These same renormalization group techniques allow one to use the TCSA to predict gaps
in actual material systems which possess a finite bandwidth/cutoff \cite{konik_scnt2}.

Below we will apply the TCSA+NRG to study the (1+1)-dimensional $SU(2)_k$ Wess-Zumino-Novikov-Witten (WZNW) 
model perturbed by the trace of the adjoint operator.  This is a strongly relevant operator with scaling dimension $d = \frac{4}{k+2}$ 
ideal for application of the TCSA+NRG method.  This perturbed conformal field theory appears in applications
such as theories of spin ladders \cite{philippe} (see also the Appendix).  A variant of this theory,
perturbing $SU(2)_k$ by the trace of the adjoint on the boundary of the system, describes
a particular class of Kondo models \cite{akhanjee}. Another variant of the model, with an additional 
current-current perturbation, 
appeared in the description of fermionic cold atoms loaded into a one-dimensional 
optical lattice \cite{coldatom1,coldatom2}.

The perturbed CFT is not integrable except at $k=2$ when it is equivalent to the theory of three massive Majorana fermions. 
Below we will use the semiclassical approximation to analyze the case of $k\gg 1$ while using TCSA+NRG for small finite values of $k$. 
Our investigations yield concrete predictions for the vacuum structure and low-energy excitations for systems described by this perturbed conformal 
field theory.

The most striking property of the theory is the dependence of its properties on the sign of the coupling constant, $\lambda$. 
For $\lambda >0$ there is a dichotomy in behavior between even and odd $k$.  For odd $k$ the semiclassical analysis predicts a massless RG flow from the $SU(2)_k$ 
critical point to $SU(2)_1$ in the infrared.  The spectrum for even $k$ is always massive and the lowest multiplet is a triplet.   However the size of the mass 
depends on the sign of $\lambda$ so that the mass is smaller for $\lambda >0$ with the ratio $m(-\lambda)/m(\lambda)$ increasing exponentially with $k$.

\section{The perturbed $SU(2)_k$ WZNW model}

The model in which we are interested is the $SU(2)_k$ WZNW model perturbed by the trace of the primary field in the adjoint representation:
\begin{eqnarray}
S&=& kW(g) + \lambda \sum_{a=1}^3\int d^2x \mbox{Tr}[\s^a g\s^a g^+]\label{action};\\
W(g) &=& \frac{1}{16\pi} \int d^2 x \; Tr(\partial^{\mu} g^{\dagger} \partial_{\mu} g) 
+ \Gamma(g) ;\nonumber \\
\Gamma(g) &=& \frac{1}{24\pi}   \int_B d^3 y \; \epsilon^{\alpha \beta \gamma} 
 Tr(g^{\dagger} \partial_{\alpha} g g^{\dagger} \partial_{\beta} g g^{\dagger} \partial_{\gamma} g),
\label{WZW}
\end{eqnarray}
where $\Gamma(g)$ is the famous Wess-Zumino term and $\s^a$ are the Pauli matrices. 
 The perturbation is equivalent to the trace of the WZNW principal field in the adjoint representation:
\be
\mbox{Tr}[\s^a g\s^b g^\dagger] \sim \Phi^{ab}_{adj},
\label{adjointfield}
\ee
This is a strongly relevant operator with scaling dimension $
d = \frac{4}{k+2}$ 
and as such it generates a characteristic energy scale 
\be
m \sim |\lambda|^{1/(2-d)}, \label{RG}
\ee
below which the spectrum is strongly modified.

It is interesting to note that the model perturbed by a single component of matrix $\Phi_{adj}$ is integrable. 
It was demonstrated in \cite{tsv87} that the perturbed Hamiltonian decomposes into a massless U(1) CFT and a massive Z$_k$ CFT 
perturbed by the thermal operator.  The properties of the latter massive theory were studied in \cite{tsv87,integrable}. 
However with the inclusion of the entire trace, integrability is lost for $k>2$.  At $k=2$ the model is equivalent to the model 
of three massive Majorana fermions with mass $m \sim \lambda$.  
In this form it has been used to describe the spin $S=1$ spin ladder \cite{tsvladder}. 

It is interesting to note that for $k=4$ when the central charge of the critical WZNW theory is equal to $c=2$, the model can be recast in abelian form:
\bea\label{twobosons}
&& S = \int d^2x\Big[\frac{1}{8\pi}\sum_{a=1,2}(\p_{\mu}\phi_a)^2 - \lambda \sum_{i=1}^3\cos\Big(e_a^{(i)}\phi_a\Big)\Big], \label{k4}\\
&& \Big({\bf e}^{(i)}\Big)^2 = 2/3, ~~ \Big({\bf e}^{(i)}{\bf e}^{(j)}\Big) = - 1/3,\nonumber
\eea
as well as the $SU(3)_1$ WZNW model perturbed by the trace of the matrix operator:
\bea
S = W[SU(3);g] + \lambda Tr(g+g^\dagger). \label{su3}
\eea

To get a qualitative understanding of the spectrum of model (Eqn. \ref{action}), we consider the case $k\gg 1$ where the model  
can be treated semiclassically. Using the identity
\be
\sum_{a=1}^3 Tr[\s^a g\s^a g^\dagger ] = 2 Trg Tr g^\dagger - 2
\ee
and the fact that the SU(2) matrix $g$ can be written as  
\be
g = n_0\hat I + \ri \s^an^a, ~~ n_0^2 + {\bf n}^2 =1,
\ee
we obtain the perturbation in the form 
\be
\sum\limits_a \mbox{Tr}[\s^a g\s^a g^\dagger] = \sum\limits_a \Phi^{aa}_{adj} = 2 n_0^2.
\ee

\begin{figure}
\begin{centering}
\psfrag{V}{$V$}
\psfrag{f}{$\Phi_1$}
\psfrag{g}{$\Phi_2$}
\includegraphics[scale=0.5]{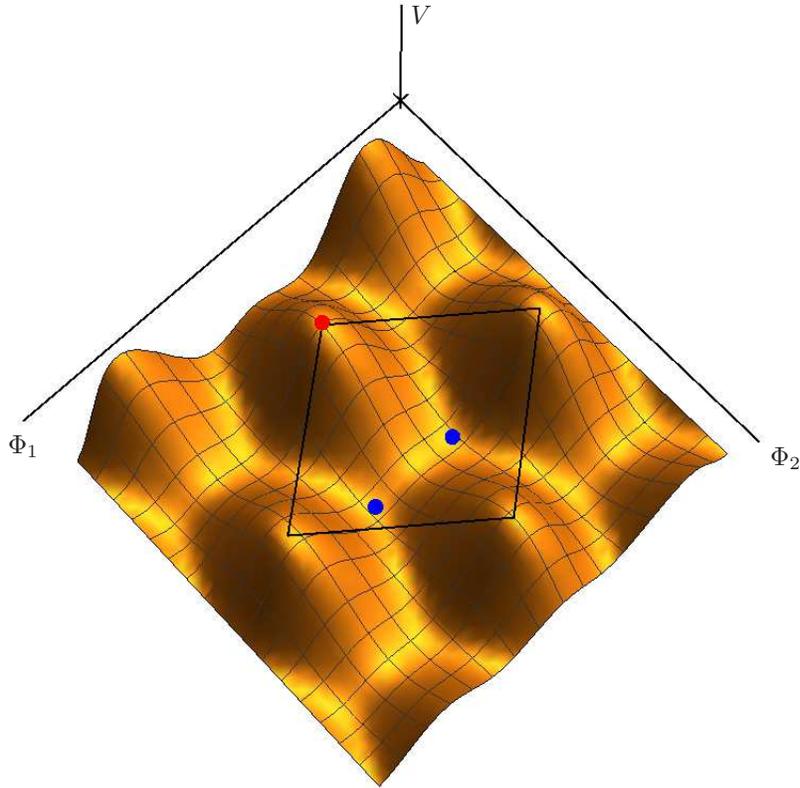}
\par\end{centering}
\protect\caption{\label{fig:semiclpot}The potential surface for $k=4$. The black
lines show the elementary cell of the periodic potential, blue dots
show the two minima relevant for $\lambda<0$ that are connected via a
saddle point. The red dot is a maximum of the potential, which becomes
the only minimum (per elementary cell) for $\lambda>0$. }
\end{figure}

For $\lambda<0$ the ground state is doubly degenerate, i.e. $n_0 = \pm 1$.
(For the SU(3) model (Eqn. \ref{su3}) this degeneracy corresponds to two possible choices of the $g$ matrix: $g = e^{\pm 2\pi i/3} I$.)
For a given choice of the sign one can consider deviations of the field $g$ from the vacuum configuration  as small. 
Then the low energy theory  becomes a theory of three weakly interacting bosons with a Lagrangian density given by
\be
{\cal L}_{eff} = \frac{k}{4\pi}(\p_{\mu}{\bf n})^2 + 2 |\lambda|{\bf n}^2 +...
\ee
where the dots stand for higher order terms. By rescaling $n^a \rightarrow n^a/k^{1/2}$ we see that these terms contain powers of $k^{-1} \ll 1$. Here the mass scale, 
\begin{equation}
m \sim \sqrt{\frac{|\lambda|}{k}},
\label{lambdapos_mass} 
\end{equation}
is obviously the one which is envisaged by RG considerations as $k\rightarrow \infty$ (see Eqn. \ref{RG}). 
The double degeneracy of the vacuum is in agreement with the structure of the potential in the abelian action (Eqn. \ref{k4}), 
which is periodic under translations with a two-dimensional lattice generated by the vectors $\mathbf{e}^{(i)}$ with two minima in each elementary
cell, as shown in Fig. \ref{fig:semiclpot}. 

For $\lambda>0$ the situation becomes more interesting.  At energies below $m$, the
field $n_0$ is suppressed and ${\bf n}$ becomes a unit vector. 
As a consequence the Wess-Zumino term (\ref{WZW}) becomes a topological one \cite{affleck,shelton}:
\be
\Gamma(\ri\s^an^a) = \frac{\ri}{8}\int d^2x \epsilon_{\mu\nu}\Big({\bf n}[\p_{\mu}{\bf n}\times\p_{\nu}{\bf n}]\Big) \equiv \ri\pi\Theta.
\ee
The coefficient of the topological term in (\ref{action}) is $\pi k$. Since the value of $\Theta$ is always integer, $\ri k\pi\Theta$ contributes  
nontrivially to the action only when $k$ is odd. In summary, the low energy effective action for $\lambda >0$ is
\bea
S = \frac{k}{4\pi}\int d^2x (\p_{\mu}{\bf n})^2 + \ri\pi k\Theta, ~~ {\bf n}^2=1.
\label{eqn:sigmaaction}\eea
This model is exactly solvable. For $k$ even \cite{zamzam,O3} the particles are massive triplets with mass 
\bea
M_{tr} \sim m k\exp(- k/2), ~~ m \sim \lambda^{1/2}. \label{scale}
\eea
The triplet structure of the particle multiplet agrees with the result for $k=2$.
However note that the mass scale at $\lambda >0$ is much smaller than the RG scale (Eqn. \ref{RG}). 
For $k$ odd, the scale $m$ (\ref{scale}) marks instead a crossover into a basin of attraction of the critical point of the $SU(2)_1$ WZNW model \cite{theta}.

\section{TCSA for the perturbed WZNW model}

\subsection{The truncated conformal space approach for $SU(2)_{k}$ perturbed by the trace
of adjoint}

For a numerical determination of the spectrum we use the truncated
conformal space approach \cite{truncated}, adapted to the $SU(2)_{k}$ 
WZNW model. For perturbations of WZNW models with levels $k=1$ and $2$ 
TCSA was applied previously in \cite{wznwtcsa}; for the present work we developed 
a general purpose TCSA code working for all $k$ and any perturbing operator.
On a Euclidean space-time cylinder of circumference, $R$, in the spatial 
direction, $x$, the Hamiltonian has the form, 
\begin{equation}
H=H_{k}+\lambda\int_{0}^{R}dx\Phi(0,x),
\label{eq:pcftham}\end{equation}
where $H_k$ is the Hamiltonian of the $SU(2)_{k}$ WZNW model (Eqn. \ref{WZW}) and the 
perturbing operator $\Phi$ is minus the trace of the adjoint
field (Eqn. \ref{adjointfield}). Due to translation invariance, the Hamiltonian is block-diagonal
on eigenspaces of the conformal spin $L_{0}-\bar{L}_{0}$. The symmetry
algebra is generated by Kac-Moody currents, $J^{\alpha}(z)$, which
satisfy the OPE,
\begin{equation}
J^{\alpha}(z)J^{\beta}(w)=\frac{k}{2}\frac{q^{\alpha\beta}}{(z-w)^{2}}+\frac{f_{\gamma}^{\alpha\beta}J^{\gamma}(w)}{z-w}+O(1),
\end{equation}
where $q^{\alpha\beta}$ is the invariant metric of the Lie-algebra
$su(2)$, and $f_{\gamma}^{\alpha\beta}$ are the structure constants.
In the basis, $\{S^{0},S^{\pm}\}$, the $su(2)$ algebra relations can be
written as
\begin{eqnarray}
[S^{\alpha},S^{\beta}] & = & f_{\gamma}^{\alpha\beta}S^{\gamma},\qquad f_{\gamma}^{\alpha\beta}=0\qquad\alpha+\beta\neq\gamma;\\
&& f_{+}^{0+} = -f_{+}^{+0}=-f_{-}^{0-}=f_{-}^{-0}=1;\qquad f_{0}^{+-}=-f_{0}^{-+}=2,\nonumber 
\end{eqnarray}
and the metric is
\begin{eqnarray}
q^{\alpha\beta} & = & 0\qquad\alpha+\beta\neq0;\\
q^{00} & = & 1;\qquad q^{\pm\mp}=2.\nonumber 
\end{eqnarray}
The modes of the current obey the Kac-Moody algebra,
\begin{eqnarray}
J_{n}^{\alpha}(z) & = & \oint_{z}\frac{d\zeta}{2\pi i}(\zeta-z)^{n}J^{\alpha}(\zeta);\\{}
[J_{n}^{\alpha},J_{m}^{\beta}] & = & f_{\gamma}^{\alpha\beta}J_{n+m}^{\gamma}+\frac{k}{2}mq^{\alpha\beta}\delta_{n+m,0},\nonumber 
\end{eqnarray}
(where $z$ denotes an arbitrary reference point of insertion; whenever
omitted, it is taken to be $z=0$). The energy-momentum tensor is
given by the Sugawara construction,
\begin{eqnarray}
T(z) & = & \frac{1}{k+2}q_{\alpha\beta}:J^{\alpha}(z)J^{\beta}(z):;\\
L_{n}(z) & = & \oint_{z}\frac{d\zeta}{2\pi i}(\zeta-z)^{n+1}T(\zeta)=\frac{q_{\alpha\beta}}{k+2}\sum_{m\in\mathbb{Z}}:J_{m}^{\alpha}J_{n-m}^{\beta}:,\nonumber 
\end{eqnarray}
where the modes $L_{n}$ satisfy the Virasoro algebra,
\begin{equation}
[L_{n},L_{m}]=L_{n+m}+\frac{c}{12}n(n^{2}-1)\delta_{n,-m},\qquad c=\frac{3k}{k+2}.
\end{equation}
We recall that for level $k$ integer there are $k-1$ primary field
multiplets $\Phi_{m,\bar{m}}^{(j)}(z,\bar{z})$, with $j=0$, $1/2$,
$\ldots$, $k/2$ and $m,\bar{m}=-j,-j+1,\dots,j$. They are normalized
as 
\begin{eqnarray}
\langle\Phi_{m_{1},\bar{m}_{1}}^{(j_{1})\dagger}(z_{1},\bar{z}_{1})\Phi_{m_{2},\bar{m}_{2}}^{(j_{2})}(z_{2},\bar{z}_{2})\rangle & = & \delta_{j_{1}j_{2}}\delta_{m_{1}m_{2}}\delta_{\bar{m}_{1}\bar{m}_{2}}(z_{12}\bar{z}_{12})^{-2h(j_{1})}\nonumber; \\
h(j) &=&\frac{j(j+1)}{k+2};\nonumber \\
\Phi_{m,\bar{m}}^{(j)\dagger}(z,\bar{z}) & = & (-1)^{2j-m-\bar{m}}\Phi_{-m,-\bar{m}}^{(j)}(z,\bar{z}).
\end{eqnarray}
The Hilbert space of the conformal field theory is given by
\begin{equation}
\mathcal{H}_{k}=\bigoplus_{j}\mathcal{V}_{j}\otimes\bar{\mathcal{V}}_{j},
\label{eq:confHilbertspace}\end{equation}
where $\mathcal{V}_{j}$ is the irreducible representation of $su(2)_{k}$
with highest weight $j$, and the bar denotes the antiholomorphic
component. The ground state level of the module $\mathcal{V}_{j}\otimes\bar{\mathcal{V}}_{j}$
is spanned by the multiplet,
\begin{equation}
|m,\bar{m}\rangle_{j}=\Phi_{m,\bar{m}}^{(j)}(0,0)|0\rangle,
\end{equation}
where $|0\rangle$ is the $SL(2,\mathbb{C})$ invariant conformal
vacuum state. The module is generated by the raising operators $J_{m}^{\alpha}$,
$m<0$; care must be taken to factor out null vectors to obtain the
irreducible representation.

We remark that for computational simplicity the left and right Kac-Moody algebras 
were implemented in the same way. This is in contrast with the usual WZNW formalism, where 
the fundamental field transforms as
\begin{equation}
 g \rightarrow g_L g g_R^{-1}.
\end{equation}
However we take our fields to transform as 
\begin{equation}
 \Phi_{m,\bar{m}}^{(j)} \rightarrow D^{(j)}(g_L)_{mm'}D^{(j)}(g_R)_{\bar{m}\bar{m}'}\Phi_{m',\bar{m}'}^{(j)} ,
\end{equation}
where $D^{(j)}$ is the SU(2) representation corresponding to spin $j$. This is related 
to the usual definition by a contragredient transformation applied to $g_R$ which is equivalent to a redefinition of the basis for SU(2).

The operator product coefficients of primary fields were derived in
\cite{fateevzamolodchikov}. Introducing the generating function fields
\begin{equation}
\Phi^{(j)}(x,\bar{x};z,\bar{z})=\sum_{m,\bar{m}}\sqrt{\left({2j\atop m+j}\right)\left({2j\atop \bar{m}+j}\right)}
x^{j-m}\bar{x}^{j-\bar{m}}\Phi_{m,\bar{m}}^{(j)}(z,\bar{z}),\label{eq:xprimary_in_components}
\end{equation}
which have the two-point function,
\begin{equation}
\left\langle \Phi^{(j_{1})}(x_{1},\bar{x}_{1};z_{1},\bar{z}_{1})\Phi^{(j_{2})}(x_{2},\bar{x}_{2};z_{2},\bar{z}_{2})\right\rangle =
\delta_{j_{1}}^{j_{2}}(x_{12}\bar{x}_{12})^{2j_{1}}(z_{12}\bar{z}_{12})^{-2h(j_{1})},
\end{equation}
the operator algebra is fully specified by the three-point functions
of the primary fields given by 
\begin{eqnarray}
 &  & \left\langle \Phi^{(j_{1})}(x_{1},\bar{x}_{1},z_{1},\bar{z}_{1})\Phi^{(j_{2})}(x_{2},\bar{x}_{2},z_{2},\bar{z}_{2})\Phi^{(j_{3})}(x_{3},\bar{x}_{3},z_{3},\bar{z}_{3})\right\rangle \nonumber \\
 &  & =C(j_{1},\, j_{2},\, j_{3})(x_{12}\bar{x}_{12})^{j_{1}+j_{2}-j_{3}}(x_{13}\bar{x}_{13})^{j_{1}+j_{3}-j_{2}}(x_{23}\bar{x}_{23})^{j_{2}+j_{3}-j_{1}}\nonumber \\
 &  & \quad\times(z_{12}\bar{z}_{12})^{h(j_{3})-h(j_{1})-h(j_{2})}(z_{13}\bar{z}_{13})^{h(j_{2})-h(j_{1})-h(j_{3})}(z_{23}\bar{z}_{23})^{h(j_{1})-h(j_{2})-h(j_{3})},
\end{eqnarray}
where the structure constants $C$ are given by
\begin{eqnarray}
C(j_{1},\, j_{2},\, j_{3})^{2} & = & 
\gamma\left(\frac{1}{k+2}\right)P(j_{1}+j_{2}+j_{3}+1)^{2}\prod_{n=1}^{3}\frac{P(j_{1}+j_{2}+j_{3}-2j_{n})^{2}}{\gamma\left(\frac{2j_{n}+1}{k+2}\right)P(2j_{n})^{2}};\nonumber \\
 &  & \gamma(x)=\frac{\Gamma(x)}{\Gamma(1-x)}\qquad P(j)=\prod_{n=1}^{j}\gamma\left(\frac{n}{k+2}\right),
\end{eqnarray}
and are fully symmetric in their arguments. Structure constants for
component fields can be obtained by expanding in the variables $x,\bar{x}$
and using Eqn. \ref{eq:xprimary_in_components}. 

The trace of the adjoint field can be expressed as the component of
the $j=1$ field which is a singlet under the global SU(2) symmetry generated by 
$J^a_0+\bar{J}^a_0$:
\begin{equation}
\Phi=\frac{1}{\sqrt{3}}\left(\Phi_{1,-1}^{(1)}+\Phi_{-1,1}^{(1)}-\Phi_{0,0}^{(1)}\right),
\label{eq:tcsapertfield}\end{equation}
where the prefactor ensures that the conformal two-point function of the perturbing field 
is canonically normalized:
\begin{equation}
\langle\Phi(z,\bar{z})\Phi(w,\bar{w})\rangle=\frac{1}{|z-w|^{4h}}, \qquad h=\frac{2}{k+2}.
\end{equation}
In fact, the field $\Phi$ as defined in Eqn. \ref{eq:tcsapertfield} differs from the trace adjoint defined previously by a sign, 
which is consistent with the form of the Hamiltonian in Eqn. \ref{eq:pcftham}. This sign can be identified from matching the 
TCSA result against the semiclassical predictions, performed in the sequel.

Using complex coordinates $\zeta=\tau+ix$, after the exponential
mapping from the plane to the cylinder,
\begin{equation}
z=e^{2\pi\zeta/R},
\end{equation}
the Hamiltonian can be written as 
\begin{eqnarray}
H & = & \frac{2\pi}{R}\left(L_{0}+\bar{L_{0}}-\frac{k}{4(k+2)}\right)+\lambda\frac{2\pi}{R}\frac{R^{2-2h}}{(2\pi)^{1-2h}}\Phi(1,1).
\end{eqnarray}
The dimensionful coupling constant $\lambda$ can be used to define a
mass scale $M$
\begin{equation}
|\lambda|=M^{2-2h}.
\label{eq:mass_scale}\end{equation}
In all our subsequent computations we use dimensionless quantities
measured in units of $M$. The dimensionless volume parameter is given
by $r=MR$ and the Hamiltonian can be written as 
\begin{equation}
\frac{H}{M}=\frac{2\pi}{r}\left[L_{0}+\bar{L_{0}}-\frac{k}{4(k+2)}+\mbox{sign}(\lambda)\frac{r^{2-2h}}{(2\pi)^{1-2h}}\Phi(1,1)\right].\label{eq:dimlessTCSAham}
\end{equation}
The matrix elements of the perturbing operator between descendant
states can be computed by a recursive procedure using the relations
given by the Kac-Moody algebra. Truncating the Hilbert space at some
descendant level $N$, the dimensionless Hamiltonian becomes a finite
numerical matrix for any given value of $r$, which can be diagonalized
numerically, resulting in a raw TCSA spectrum that depends on the
truncation level. In many cases the raw TCSA data already gives an
accurate spectrum; however, we used both numerical and analytic schemes
to eliminate cut-off dependence and obtain better results. 

Since the perturbation is a singlet, it conserves the $z$ component of the diagonal SU(2) ($g_L=g_R$) which is 
\begin{equation}
\mathcal{Q}=J^z_0+\bar{J}^z_0.
\label{eq:Sz_charge}
\end{equation}
All eigenvalues of the charge $\mathcal{Q}$ are integers and the Hilbert space can be decomposed into sectors labeled by 
the eigenvalues of $\mathcal{Q}$. In all our computations we only show results for states with $\mathcal{Q}=0$. Since the 
Hilbert space decomposes into integer spin representations of the diagonal SU(2), each such multiplet has a single level 
lying in the $\mathcal{Q}=0$ subspace. It is guaranteed analytically, but we also checked numerically that any 
$\mathcal{Q}\neq 0$ state is degenerate with one of the levels in the $\mathcal{Q}=0$ subspace, and that degenerate 
states form full multiplets. In addition, all TCSA data we present correspond to zero-momentum states (i.e. $L_0-\bar{L}_0=0$), 
as non-zero momentum subspaces contain no new physics.

It is also important to observe that according to the Kac-Moody fusion rules under the perturbation (Eqn. \ref{eq:tcsapertfield}),
the Hilbert space (Eqn. \ref{eq:confHilbertspace}) decomposes into even and odd sectors which originate from states with $j$ integer and 
half-integer, respectively. This will be important in the sequel.

\subsection{Testing the TCSA}

In this subsection we describe a number of tests that we put our TCSA code through.  Such tests are
very important as there are no exact
results to verify the TCSA due to the non-integrability of the theory.  The first test that we
considered is specific to $k=4$.   As we have discussed, we have two realizations of $SU(2)_4$+${\rm Tr}\Phi_{adj}$:
one treating the conformal basis of $SU(2)_4$ in the language of current algebras (the picture we are focusing on in this paper) and one treating this same
basis as a two boson theory, one boson with compactification radius of $\sqrt{2}\pi$ and one orbifolded boson
with compactification radius of $2\sqrt{6}\pi$ (see Eqn. \ref{twobosons}).  While these are very different starting points for the TCSA, they must lead
to the same answer.  And we have checked, at least in the sector of the theory containing the ground state, that they do.  
While this is an important check of the code, it only applies to $k=4$. 
We have thus also considered the case $SU(2)_{1}$
perturbed with the singlet component of the $j=1/2$ primary field
(equivalent to sine-Gordon theory in the SU(2) symmetric point
of the attractive regime) as well as the $k=2$ case, where the spectrum
must agree with that of three Majorana fermions; some details on this
latter case are given in subsection \ref{subsec:k2}.  Both of these
tests indicate the code is working.  Finally, as discussed in subsection \ref{subsec:rgs},
we show that corrections to the ground state due to changing the cutoff
in the theory as determined by the TCSA match those computed {\it analytically} using conformal
field theory.   This analytical computation is non-trivial and so provides another important check on
whether the code is behaving as it should.

\subsection{Renormalization methods}

The first improvement to the raw TCSA is given by the 
numerical renormalization
group (NRG) method introduced in \cite{tsrg}. The procedure
consists of starting at a cut-off value where the whole matrix can
be diagonalized and then incorporating higher energy levels in chunks
of a given step size (number of states added at each step) until the
target value of the cut-off is reached. This is necessary as the number
of states grows very fast with the cut-off. For example, in the integer
$j$ sector of the $k=4$ theory for descendant levels $N=3,4,5,6,7$ we have $1427$,
$6373$, $23498$, $83144$ and $264129$ zero-momentum states respectively in the 
$\mathcal{Q}=0$ sector. Our computational
capacity allowed us to reach the descendant level $N=5$ with exact diagonalization, while
to reach $N=6,7$ we have used the NRG procedure. 

The next improvement takes into account the contribution of states
above the cut-off using perturbation theory to second order in $\lambda$.
There are several schemes in the literature that can be used, but
a compromise must be struck between computational costs and accuracy.
Below we give a short description of each procedure, and compare them
in order to make an optimal choice for our problem.

\subsubsection{Vacuum counter term}

The contribution from the omitted high-energy states dependence of the 
ground state energy was computed using the results in \cite{takacs}, 
and the counter term necessary to eliminate the cut-off dependence to 
second order is given by  
\begin{eqnarray}
\delta E_{0} & = & \frac{\pi R^{3-4h}}{2(2\pi)^{2-4h}}\frac{1}{h+N+1}\left(\frac{\Gamma(2h+N+1)}{\Gamma(2h)\Gamma(N+2)}\right)^{2}\nonumber \\
 &  & \times\phantom{}_{4}F_{3}(1,1+h+N,1+2h+N,1+2h+N;2+N,2+N,2+h+N;1)\nonumber \\
 & = & \frac{(2\pi)^{4h-1}R^{3-4h}}{4(2h-1)\Gamma(2h)^{2}}N^{4h-2}+\ldots\label{eq:vac_counterterm}
\end{eqnarray}
with $\phantom{}_{4}F_{3}$ denoting a generalized hypergeometric
function.

\subsubsection{Counter terms for excited states}

To eliminate cut-off dependence for excited states we can use a scheme
developed in \cite{rychkov}. The idea is to separate
the Hilbert space into a low-energy part (labeled by $l$), which
is included in TCSA, and a high-energy part (labeled by $h$) which
consists of states above the cut-off. For any state we split its eigenvector
$c$ into low- and high-energy parts $c_{l}$ and $c_{h}$; similarly,
the Hamiltonian can be split into a block form according to
\begin{equation}
H=\left(\begin{array}{cc}
H_{ll} & H_{lh}\\
H_{hl} & H_{hh}
\end{array}\right).
\end{equation}
The full eigenvalue problem can be split accordingly 
\begin{equation}
H_{ll}c_{l}+H_{lh}c_{h}=\varepsilon c_{l},\quad H_{hl}c_{l}+H_{hh}c_{h}=\varepsilon c_{h},
\end{equation}
where $\varepsilon$ is the exact eigenvalue. Eliminating the high-energy
components, $c_{h}$, gives 
\begin{equation}
[H_{ll}\underbrace{-H_{lh}\left(H_{hh}-\varepsilon\right)^{-1}H_{hl}}_{\Delta H_{\text{full}}}]c_{l}=\varepsilon c_{l}.
\end{equation}
We write 
\begin{equation}
H=H_{0}+V,
\end{equation}
where $H_{0}$ is the conformal Hamiltonian and $V$ is the matrix
of the perturbing field. Note that the off-diagonal components only
involve the perturbing matrix, so we obtain 
\begin{eqnarray}
\Delta H_{\text{full}} & = & -V_{lh}(H_{0}+V_{hh}-\varepsilon)^{-1}V_{hl}\nonumber \\
 & \approx & \underbrace{-V_{lh}(H_{0}-\varepsilon)^{-1}V_{hl}}_{\Delta H}+O(E_{max}^{-2}),\label{eq:deltaH_approx}
\end{eqnarray}
where $E_{max}$ is the cut-off in energy units, and we used the fact
that $V_{hh}$ only contributes at higher order. First order perturbation
theory in $E_{max}^{-1}$ gives 
\begin{equation}
\varepsilon=E_{TCSA}+c_{TCSA}\Delta Hc_{TCSA},\label{eq:full_Rychkov_CT}
\end{equation}
where 
\begin{equation}
H_{ll}c_{TCSA}=E_{TCSA}c_{TCSA}.
\end{equation}
To calculate $\Delta H$ we can approximate $\varepsilon\to E_{TCSA}$
in Eqn. \ref{eq:deltaH_approx} 
\begin{equation}
\Delta H_{ab}=-\int_{E_{max}}^{\infty}\frac{M(E)_{ab}}{E-E_{TCSA}},
\label{eq:deltaHab}\end{equation}
where $M(E)_{ab}$ is defined by
\begin{equation}
\langle a\vert V(\tau)V(0)\vert b\rangle=\int_{0}^{\infty}dEe^{-(E-E_{a})\tau}M(E)_{ab}.
\label{eq:MEab}\end{equation}
Here $V(\tau)$ is understood as a Euclidean time-evolved version
of the perturbing operator, 
\begin{equation}
V(0)=\int_{0}^{R}dx\Phi(ix).
\end{equation}
Then one can write 
\begin{equation}
V(\tau)V(0)=\int_{0}^{R}dx_{1}\int_{0}^{R}dx_{2}\Phi(ix_{1}+\tau)\Phi(ix_{2}),
\end{equation}
where $\Phi$ has scaling dimensions, $(h,h)$. The leading contributions
can be computed by considering the most singular terms in the operator
product expansion on the cylinder 
\begin{equation}
\Phi(ix_{1}+\tau)\Phi(ix_{2})\approx\sum_{\varphi}C_{\varphi\Phi\Phi}\left|ix_{1}-ix_{2}+\tau\right|^{-4h+2h_{\varphi}}\varphi(ix_{2})+\ldots
\end{equation}
where $\varphi$ has scaling dimensions $(h_{\varphi},h_{\varphi})$
and we only consider the first few operators. Indeed, the leading
contribution is the identity with $h_{1}=0$ and $C_{1\Phi\Phi}=1$;
since we neglect the subleading terms of the identity contribution,
we cannot include any operators with $h_{\varphi}\geq1/2$.

In the spin $0$ sector, using translation invariance of the states $\vert a\rangle$
and $\vert b\rangle$ gives 
\begin{eqnarray}
\langle a\vert\varphi(ix_{2})\vert b\rangle & = & \langle a\vert\varphi(0)\vert b\rangle\\
 & = & \left(\frac{2\pi}{R}\right)^{2h_{\varphi}}\langle a\vert\varphi_{plane}(1)\vert b\rangle.\nonumber 
\end{eqnarray}
The integrals can be explicitly computed 
\begin{eqnarray}
\int_{0}^{R}dx_{1}\int_{0}^{R}dx_{2}\left|ix_{1}-ix_{2}+\tau\right|^{\alpha} & = & 2R^{2}\tau^{\alpha}\,_{2}F_{1}\left(\frac{1}{2},-\frac{\alpha}{2};\frac{3}{2};-\frac{R^{2}}{\tau^{2}}\right)+\frac{\tau^{2\alpha+2}}{\alpha+1}-\frac{\left(R^{2}+\tau^{2}\right)^{\alpha+1}}{\alpha+1},\nonumber \\
 & = & \sqrt{\pi}R\frac{\Gamma\left(-\frac{\alpha+1}{2}\right)}{\Gamma\left(-\frac{\alpha}{2}\right)}\tau^{1+\alpha}+\text{less singular}.
\end{eqnarray}
To leading order we can neglect the term $e^{E_{a}\tau}$ in the correlator,
since the energies $E$ we consider are above the cut-off $E_{max}$,
while $E_{a}$ is below it; similarly we can take $E_{TCSA}\to 0$
in Eqn. \ref{eq:deltaHab}. This gives 
\begin{equation}
\Delta H_{ab}^{\varphi}=\frac{(2\pi)^{4h-1}R^{3-4h}N^{4h-2h_{\varphi}-2}}{4(2h-2h_{\varphi}-1)\Gamma(2h-2h_{\varphi})^{2}}
C_{\varphi\Phi\Phi}\langle a\vert\varphi_{plane}(1)\vert b\rangle ,\label{eq:leading_Rychkov_CT}
\end{equation}
for the contribution coming from the operator $\varphi$ in the OPE. Note that this equation
gives the counter term in an  operator form, in an arbitrary basis of states $|a\rangle$. 
The eventual correction to any given energy level is computed by evaluating its matrix in the 
TCSA basis and computing its expectation value with the TCSA eigenvectors corresponding 
to the given level at cutoff $N$. Note that when applied to the case of the identity, 
this result reproduces the leading behaviour of the vacuum counter term (Eqn. \ref{eq:vac_counterterm}).

\subsubsection{Running coupling}

To the leading order, the counter terms (Eqn. \ref{eq:leading_Rychkov_CT})
are independent of the state under consideration
and can be considered as local operators
added to the Hamiltonian. This forms the basis of a renormalization
group (RG) by defining a level dependent running coupling such that
the counter term describing the cut-off dependence is compensated
by changing the coupling. Let us denote

\begin{equation}
\mu=\left(\frac{2\pi}{r}\right)^{2h-2},\label{eq:mu_r}
\end{equation}
and consider the case when $\varphi=\Phi$, i.e. $h_{\varphi}=h$.
Then we can introduce a cut-off dependent coupling by requiring that
the contribution of high-energy states from level $N$ is compensated
by the change in the coupling. The contribution from the $N$th level
to the $\Phi$ term in the Hamiltonian is just the derivative of Eqn. \ref{eq:leading_Rychkov_CT}
with respect to $N$ (to leading order in $1/N$), which is equal
to 
\begin{equation}
\Delta H_{ab}^{\Phi}=\frac{2\pi}{r}\mu^{2}\frac{\pi C_{\Phi\Phi\Phi}}{\Gamma(h)^{2}}N^{2h-3}\langle a\vert\Phi_{plane}(1)\vert b\rangle.
\end{equation}
We make the coupling depend on $N$ and stipulate that it is evolved
from $N$ to $N-1$ by including the counter term's contribution.
This leads to the RG equation of the form used in \cite{watts_arxiv,watts,konik_scnt1}
\begin{equation}
\mu_{N}=\mu_{N-1}+\mu_{N-1}^{2}\frac{\pi C_{\Phi\Phi\Phi}}{\Gamma(h)^{2}}N^{2h-3}+\dots
\end{equation}
where the dots denote terms subleading for large $N$, which can be
integrated to 
\begin{equation}
\mu_{\infty}=\frac{\mu_{N}}{1+C_{1}\mu_{N}N^{2h-2}},\qquad C_{1}=\frac{2\pi C_{\Phi\Phi\Phi}}{(4h-4)\Gamma(h)^{2}}\label{eq:running_coupling},
\end{equation}
to leading order for large $N$. Let us denote the dimensionless energy
levels by $e_{i}(r)$, which are just the eigenvalues of the dimensionless
TCSA Hamiltonian (Eqn. \ref{eq:dimlessTCSAham}) as functions of $r$,
with the vacuum being $e_{0}(r)$.

This RG equation, inasmuch as it can be derived from the invariance of the partition function under changes
in coupling \cite{watts_arxiv}, expresses the RG invariance of the gaps of the form:
\begin{equation}
E_i(\lambda_\infty) = E^{(N)}_i(\lambda_N).
\end{equation}
Due to Eqn. \ref{eq:mu_r}, this invariance can be reinterpreted in 
terms of the dimensionless energy levels via
\begin{equation}
e_{i}(r_{\infty})=\frac{r_{N}}{r_{\infty}}e_{i}^{(N)}(r_{N}),\label{eq:GW_energy_levels}
\end{equation}
where $e_{i}$ are the energy levels at cut-off $N=\infty$. Note
that the energy also needs to be rescaled due to the $1/r$ prefactor
in the dimensionless TCSA Hamiltonian (Eqn. \ref{eq:dimlessTCSAham}).

We remark that at higher orders in $1/N$ the counter terms (and therefore 
the running coupling as well) are state dependent \cite{watts}. 
One way to take this into account is to compute the full
counter term (Eqn. \ref{eq:full_Rychkov_CT}) without using the approximation
of the previous subsection, i.e. keeping the dependence on $E_{TCSA}$
and $E_{a}$ in Eqns. \ref{eq:deltaHab}  and \ref{eq:MEab}. This leads to a 
rather complicated and computationally
expensive method even for the counter terms themselves, and it makes 
rather difficult the implementation and solution of the corresponding renormalization
group equations, which describe non-local Hamiltonian terms \cite{rychkov}.
Henceforth we neglect these higher corrections in our computations.

\subsubsection{\label{subsec:rgs}Renormalizing the ground state}

\begin{figure}
\protect\caption{\label{fig:bulk_energy_renormalization_with_rg}$k=3$ (top) and $4$ (bottom), 
$\lambda>0$ ground state energy data obtained by TCSA+NRG for cutoff 
levels $N=5,6,7,8$ and $N=4,5,6,7$ (dashed lines); the results with the
counter term (Eqn. \ref{eq:vac_counterterm}) (red lines); and the data
involving the counter term and the RG improvement (Eqn. \ref{eq:GW_energy_levels})
(green lines). 
The insets show the same results blown up on the $1<L<14$
interval. Note that in the direction of increasing cut-off $N$ the
subtracted (red), and the subtracted and renormalized (green) levels
move less as the cut-off grows, which is a further confirmation of
the validity of the renormalized TCSA.}
\centering{}
\psfrag{E}{$E/M$}
\psfrag{L}{$MR$}
\includegraphics{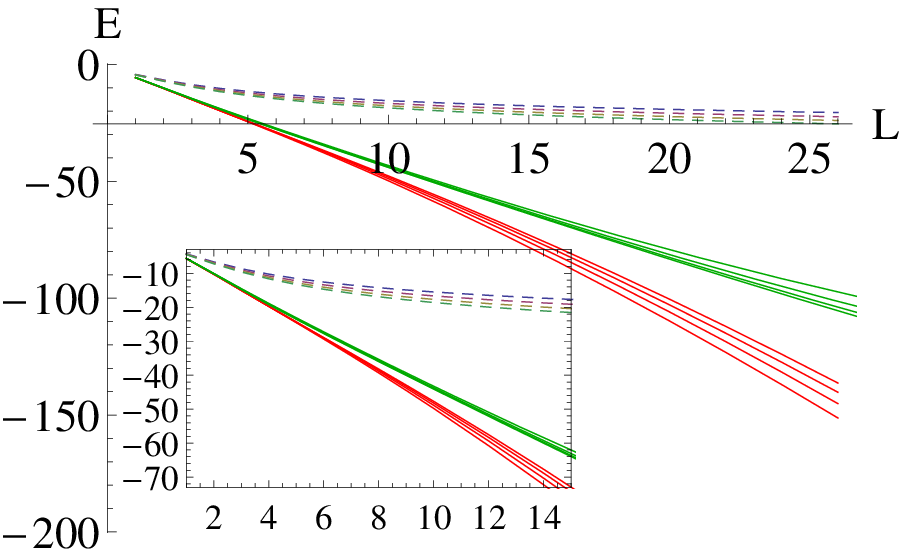} 
\\
$k=3$\\
\includegraphics{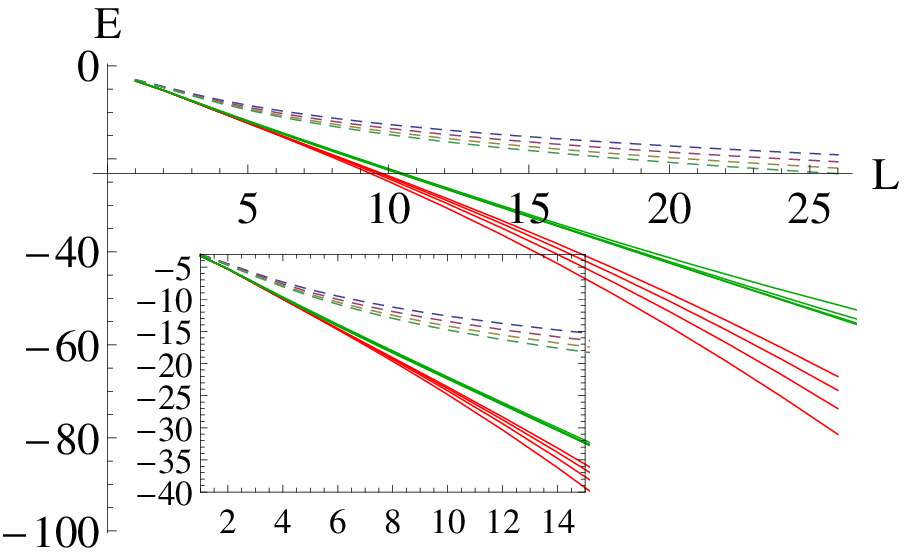}
\\
$k=4$
\end{figure}

The ground state of the theory has the conformal vacuum as its ultraviolet limit, 
and is contained in the even, $\mathcal{Q}=0$, zero-momentum sector.
In Fig. \ref{fig:bulk_energy_renormalization_with_rg} we show results
coming from NRG+TCSA, the effect of the vacuum counter term 
(Eqn. \ref{eq:vac_counterterm}), 
as well as the results obtained by implementing both the counter term 
and the running coupling according to Eqn. \ref{eq:GW_energy_levels}. 
In small volume we can see that the counter terms at different 
cut-off levels scale the 
energy level to the same curve, verifying that the subtraction 
provides reliable results even when starting from NRG+TCSA data
with low values of the cut-off. Note that taking into account 
the running coupling gives a further significant reduction 
of cut-off dependence.   This scaling is one of the verification tools
we used to affirm our belief that our code is producing correct results.

The slope extracted from the linear regime of the vacuum level gives the 
ground state (bulk) energy density $\mathcal{E}_0$, and can be estimated by fitting a 
linear function in the appropriate range of volume. The resulting estimates 
are given in Table \ref{tab:bulk}. 

\begin{table}
\begin{tabular}{|l|l|l|}
\hline
$\mathcal{E}_0$ & $\lambda>0$ & $\lambda<0$ \cr
\hline
$k=3$ & -4.55 $\pm$ 0.01 & -7.69 $\pm$ 0.01 \cr
\hline
$k=4$ & -2.21 $\pm$ 0.01 & -5.03 $\pm$ 0.01 \cr 
\hline
\end{tabular} 
\caption{\label{tab:bulk}
The bulk energy density $\mathcal{E}_0$ given in units of $M$ (cf. Eqn. \ref{eq:mass_scale}).}
\end{table}

\subsubsection{Numerical application of the renormalization methods}

To eliminate the additive bulk energy renormalization (Eqn. \ref{eq:vac_counterterm})
we consider the gaps relative to the vacuum. The running coupling
(Eqn. \ref{eq:running_coupling}) leads to the renormalization prescription
\begin{equation}
e_{i}(r_{\infty})-e_{0}(r_{\infty})=\frac{r_{N}}{r_{\infty}}\left(e_{i}^{(N)}(r_{N})-e_{0}^{(N)}(r_{N})\right),\label{eq:GW_rel_energy_levels}
\end{equation}
which we call the RC (running coupling) correction. It is also possible
instead to add the counter terms (Eqn. \ref{eq:leading_Rychkov_CT}) where
we can include the contribution of all operators $\varphi$ below
a certain $h_{\varphi}$ chosen to keep the slowest-decaying contributions 
(in practice we chose to incorporate the primary contributions). 
This will be called the CT (counter term)
correction. The difference between the two corrections is that in
contrast to the CT correction, the RC correction only involves a single
operator contribution, however by introducing the running coupling
it sums up the leading power in the cut-off dependence to any order. 

\begin{figure}[h]
\begin{centering}
\psfrag{E}{$E/M$}
\psfrag{L}{$MR$}
\includegraphics[scale=0.8]{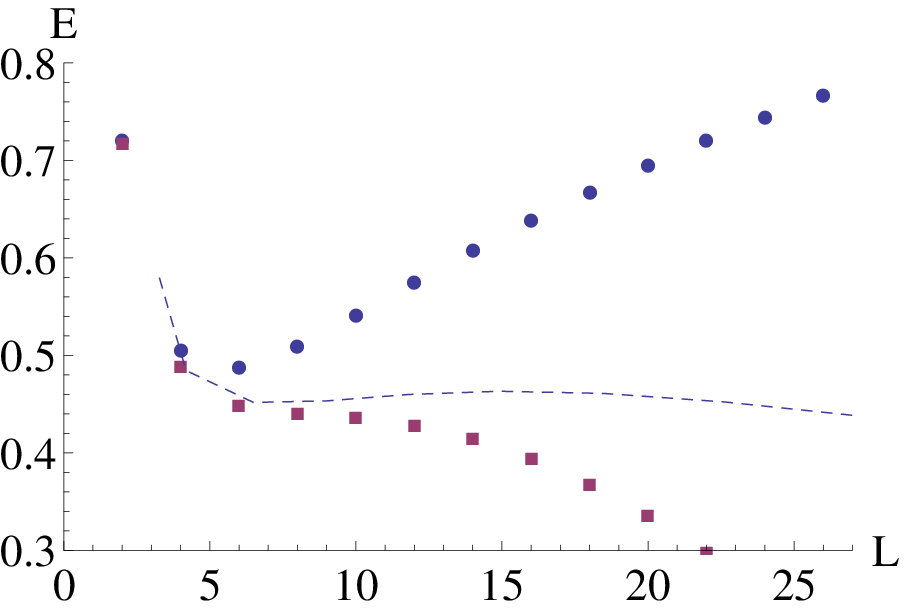}~~~~~\includegraphics[scale=0.8]{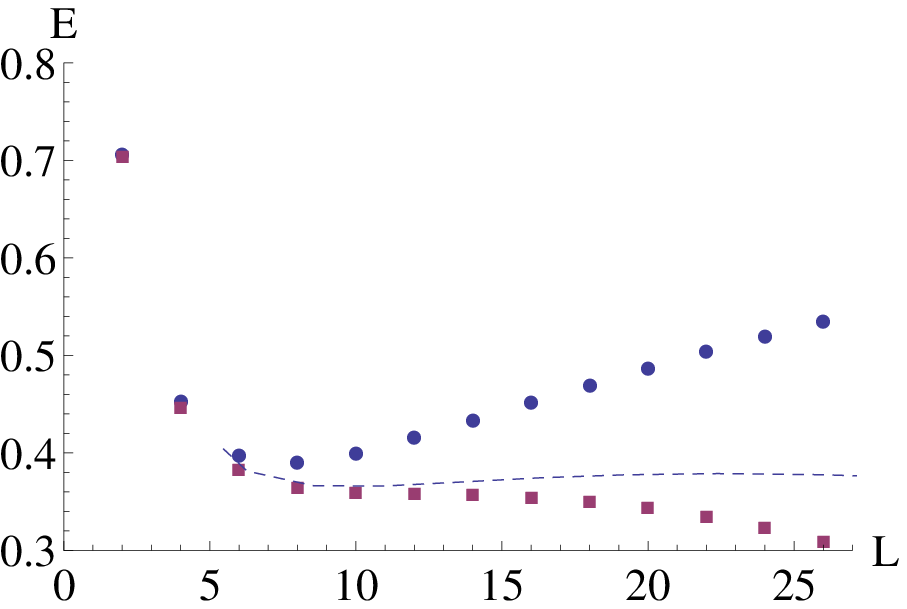}
\par\end{centering}
\begin{centering}
\psfrag{E}{$E/M$}
\psfrag{L}{$MR$}
\includegraphics[scale=0.8]{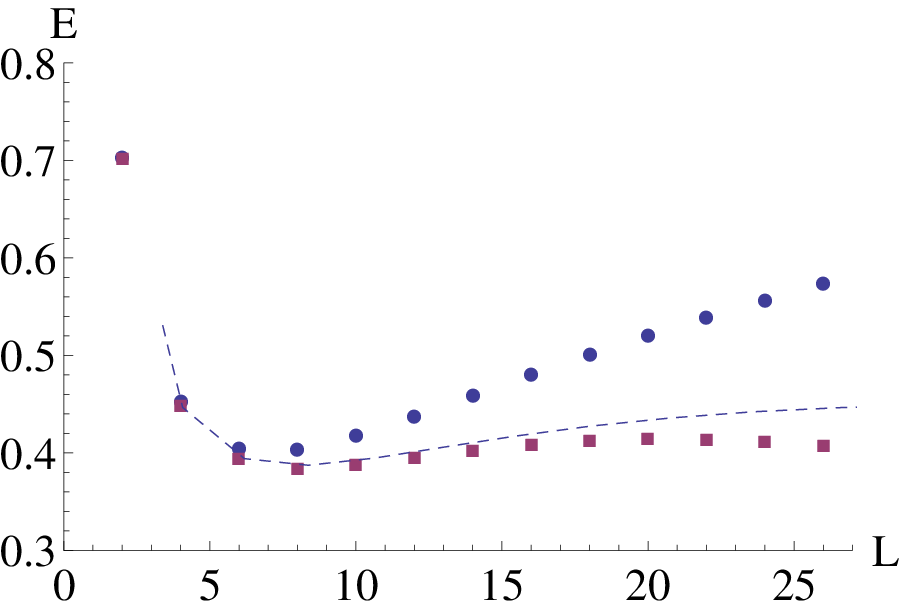}~~~~~\includegraphics[scale=0.8]{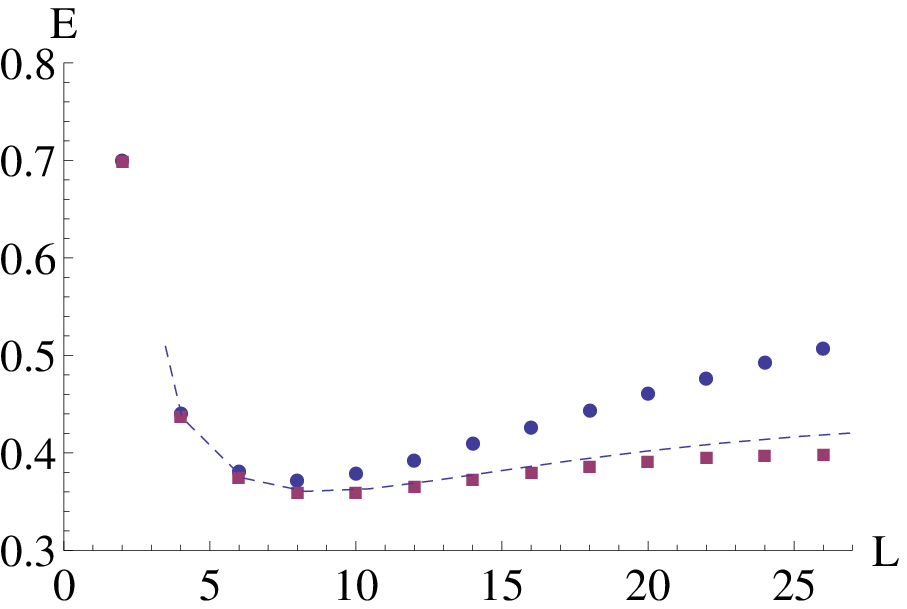} 
\par\end{centering}
\protect\protect\caption{\label{fig:GWvsRychkov}
Finite volume gap in the perturbed $SU(2)_{4}$
model for $\lambda>0$ at cut-offs $N=2$, $3$, $4$ and $5$. 
The plot shows the raw TCSA data (without NRG, blue circles), 
and those after the RC (dashed lines)
and the CT (magenta squares) corrections were applied. 
The gap can be estimated to be $\Delta=0.36\pm0.03$, 
and the error is approximated by the difference between 
the gap estimates for $N=3$ and 6.}
\end{figure}

We illustrate the renormalization method for the first excited level for $k=4$ and $\lambda>0$, 
which comes from the odd sector and consists of three degenerate states with 
$\mathcal{Q}=+1,0,-1$ forming a triplet under diagonal $SU(2)$. We consider the state
corresponding to a stationary particle, which can be found in the zero-momentum sector.
For higher cut-offs we find that the two prescriptions converge to
each other as illustrated in Fig. \ref{fig:GWvsRychkov}, so we can
use the computationally simpler RC method to obtain renormalized results.

\section{Numerical results for the spectrum}

\begin{figure}
\begin{centering}
\psfrag{E}{$E/M$}
\psfrag{L}{$MR$}
\includegraphics{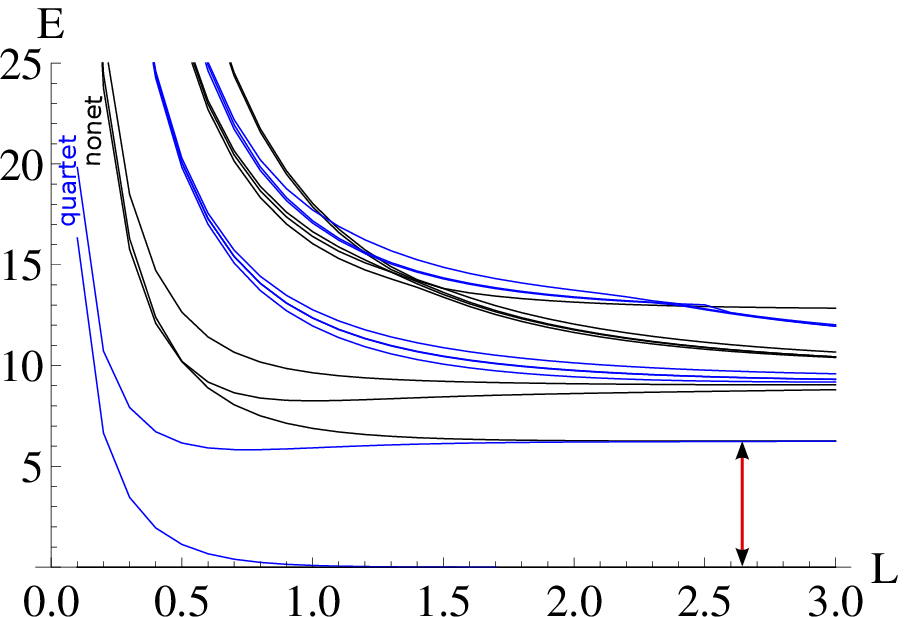}\\
\includegraphics{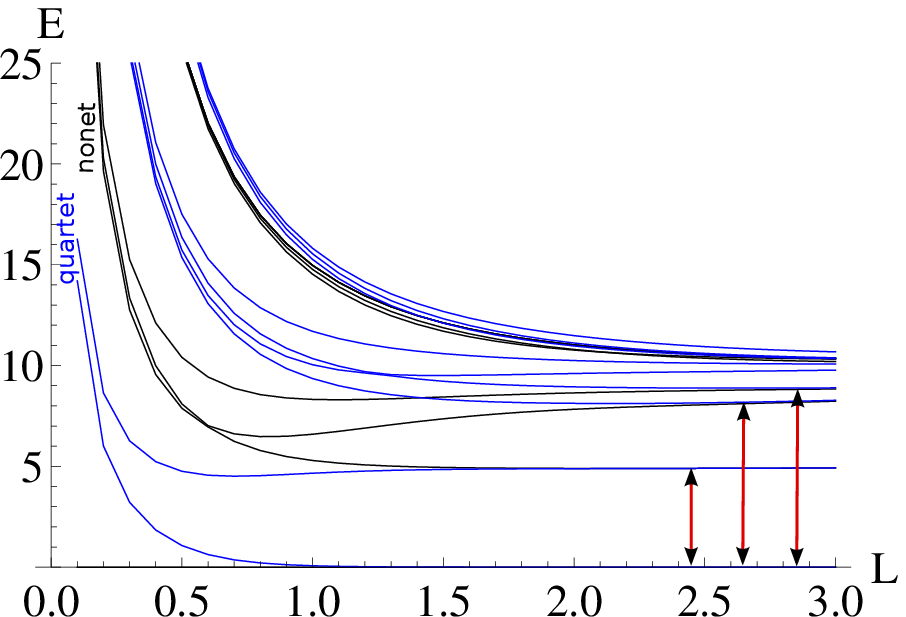}\\
\includegraphics{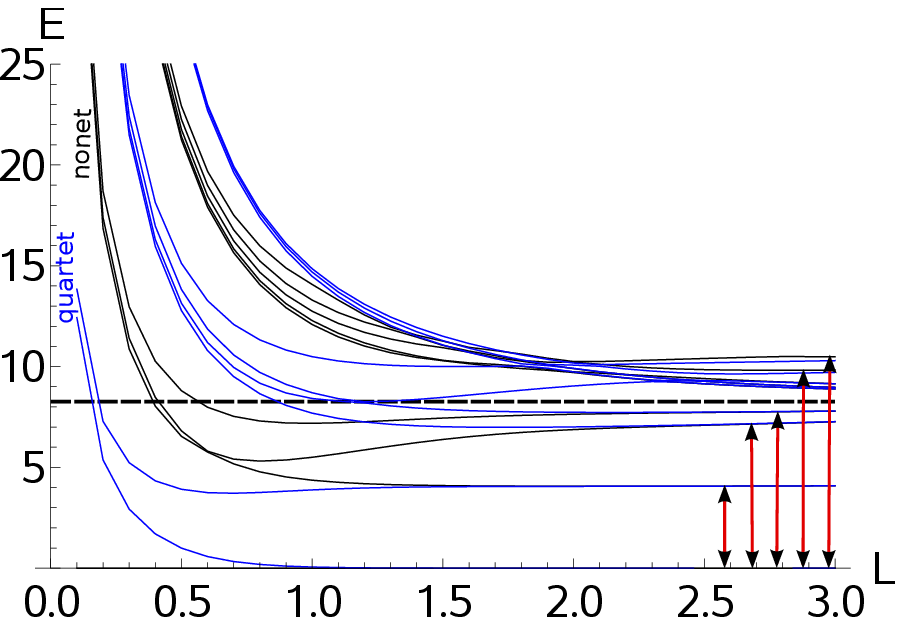} 
\par\end{centering}

\protect\caption{\label{fig:lambdaneg}Finite volume spectra in the perturbed $SU(2)_{k}$
models with $k=3,4,5$ for negative coupling constant at cut-offs
$n_{max}=7$, $6$, $5$, respectively. We show raw TCSA data since
renormalization here had an effect that is not visible on these figures. 
Colors represent energy levels in the integer
(black) and half integer (blue) sectors. The red arrows show the gaps corresponding to one-particle states. 
For $k=5$, two of these are already higher than the two-particle threshold (shown as the thick dashed line), 
and due to non-integrability they are expected to correspond to resonances.}
\end{figure}

\subsection{The $\lambda<0$ case}

In this regime we expect two triplets of  weakly interacting particles
that have a mass given by  Eqn. \ref{lambdapos_mass}. 
We show the spectra of zero-momentum $\mathcal{Q}=0$ states relative to the absolute 
ground state in Fig. \ref{fig:lambdaneg} which clearly show a doubly degenerate vacuum structure. 
In finite volume, the degeneracy of the vacua is lifted by the tunneling, which vanishes exponentially 
with the volume. Due to the $\mathbb{Z}_2$ symmetry relating the two vacua
$|1\rangle$ and $|2\rangle$, the finite volume ground states are given by 
\begin{equation}
 |\pm\rangle=\frac{1}{\sqrt{2}}\left(|1\rangle\pm|2\rangle\right),
\end{equation}
and are expected to emerge from the even and odd sectors, respectively. Since we plot the 
energies relative to the absolute finite volume ground state $|+\rangle$, the presence 
of these vacua is signalled by a state originating from the odd sector with a relative energy 
approaching zero exponentially with the volume.

As expected from the semiclassical considerations, the first excitations are indeed a triplet of particles, 
and they too appear in two copies according to the vacua. Their triplet nature can be seen 
both by looking in the $\mathcal{Q}=\pm 1$ sectors for the other components 
of the multiplets, but also from the fact that the energy levels in the ultraviolet 
($mR\sim 0$) limit are seen to emerge from conformal states transforming as a triplet
under the diagonal SU(2). In particular, the lowest lying states come from a 
quartet of states created by the primary field $\Phi^{(1/2)}$ and a nonet of states 
created by the primary field $\Phi^{(1)}$. Under the global SU(2), the quartet decomposes as
\begin{equation}
\frac{1}{2} \otimes \frac{1}{2}= 0 \oplus 1,
\label{eq:hhdecompose}\end{equation}
with the singlet giving the second vacuum state, while the triplet corresponds to the first 
triplet of one-particle states. In the plots of Fig. \ref{fig:lambdaneg} the quartet 
corresponds to the first two blue lines (their colour indicates that they come from a sector
created by a primary field with half-integer $j$).  

\begin{figure}
\begin{centering}
\psfrag{D}{$\Delta$}
\psfrag{K}{$1/\sqrt{k}$}
\includegraphics{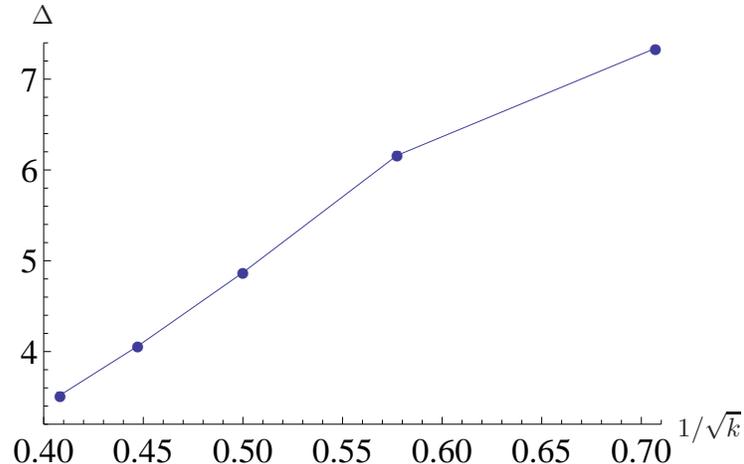}
\par\end{centering}

\protect\caption{\label{fig:TCSA-mass-gap-negative-side}TCSA mass gap for negative
coupling constant as a function of $k^{-1/2}$. We show data coming
from $k=2,$ $3$, $4$ and $5$ models. We also put error bars on
the data points which we calculated by subtracting the gap estimates
with and without applying the RC improvement, but they are so small
that they are practically invisible in the plot.}
\end{figure}

The nonet
\begin{equation}
1 \otimes 1= 0 \oplus 1 \oplus 2,
\label{eq:11decompose}\end{equation}
contains the other triplet of one-particle levels while the single and quintet are excited states. They correspond
to the lowest three black levels visible in the plots, which are exactly 
(the $\mathcal{Q}=0$ components of) the three multiplets. Of these three levels, the one corresponding to the 
triplet $1$ in Eqn. \ref{eq:11decompose} approaches the other (blue) triplet from Eqn. \ref{eq:hhdecompose} 
exponentially with increasing volume. They also level out exponentially, signaling that these 
are single-particle states, coming in two copies according to the degenerate vacua.

In Fig. \ref{fig:TCSA-mass-gap-negative-side} we show that for larger values
of $k$ the gap measured from the flat portion of the first one-particle levels indeed follows 
the $k^{-1/2}$ scaling of the particle mass expected from the semiclassical arguments.
The spectra also show the presence of additional states below the two-particle threshold. Due 
to the double degenerate vacua, one expects kinks interpolating between them. With the periodic 
boundary conditions imposed by TCSA one can only see states with an even number of kinks such that
the sequence of vacua interpolated by them has the same starting and end points. In addition, these
kinks are also expected to have bound states, and the lowest lying particle triplet can be identified 
with the lowest mass kink-antikink bound states.

In the absence of more detailed knowledge about the theory, at present
we cannot identify the higher states conclusively, but it seems that at least the first few levels 
very much resemble  the breather doublets seen in the $2$-folded sine-Gordon theory \cite{ksg}, 
so it is likely that these are indeed higher kink-antikink bound states beyond the lowest triplet.
These levels can be seen in Fig. \ref{fig:lambdaneg} as pairs of black and blue lines approaching 
each other and also leveling out exponentially at the same time. 
The fact that one of these always comes from the even, while the other from the odd sector 
(as shown by their colours) confirms the interpretation that these are indeed two copies of higher 
kink-antikink bound state multiplets (note that these are not necessary triplets; from the 
identification of the nonet lines we know the next two are a singlet and a quintuplet).

One can also see that the number of such one-particle level candidates increases with $k$, which is
what is expected in the semiclassical limit \cite{dhn,mussardo}. In addition, the characteristic dependence of 
the lowest particle mass on $k$ suggests that the mass scale $M$ is related to the kink mass, and that the 
spectrum of bound states becomes dense for large $k$, analogously to the $\Phi^4$ and sine-Gordon models 
treated in \cite{dhn,mussardo}. Due to non-integrability of the model it is also expected that two-kink bound 
states over the two-particle threshold are in fact resonances whose finite volume signatures 
must resemble those in the two-frequency sine-Gordon theory studied in \cite{resonances}; however, 
our data do not allow a reliable identification of these signatures at present. We also remark
that in spite of non-integrability, there are also apparent level crossings in the spectra, e.g. 
between even (black) and odd (blue) levels since they do not mix under the perturbation.
The additional higher states can be interpreted as two- and more particle levels composed of 
particles and/or even number of kinks.

\subsection{$\lambda>0$, even $k$}

\begin{figure}
\begin{centering}
\psfrag{E}{$E/M$}
\psfrag{L}{$MR$}
\includegraphics{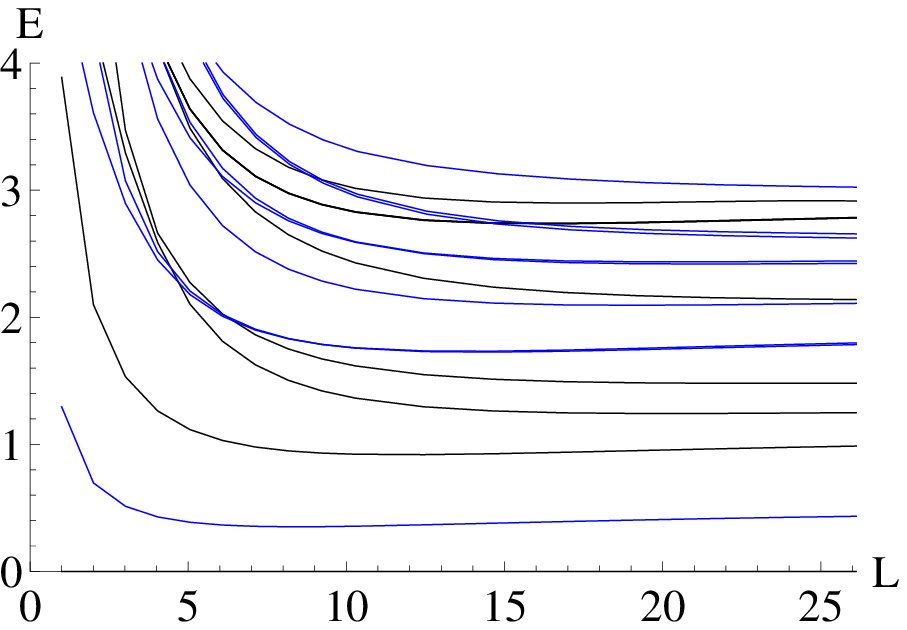} 
\includegraphics{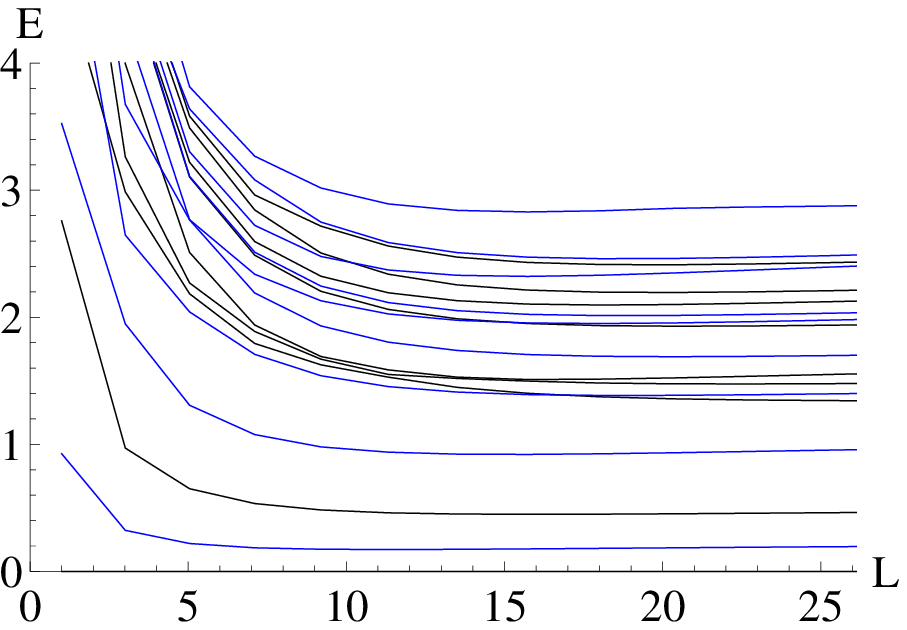}
\par\end{centering}

\protect\caption{\label{fig:TCSAposeven}Spectra in the $k=4$ and $6$ 
models for $\lambda>0$ at cut-offs $N=7$ and $4$. Data was improved
by the RC method. Colors represent level data obtained in the integer
(black) and half integer (blue) sectors.}
\end{figure}

In Fig. \ref{fig:TCSAposeven} we show the results for positive $\lambda$
and even level $k$. We observe a single vacuum and a triplet of one-particle
levels, which are consistent with the effective $\sigma$-model picture based 
on the action in Eqn. \ref{eqn:sigmaaction}.
From the first excited levels alone, the gaps can be estimated as 
$\Delta=0.37\pm0.03$ and $\Delta=0.18\pm0.02$ for $k=4$ and $6$, respectively, 
which are much smaller than those for $\lambda<0$ and decrease strongly with 
increasing $k$ as expected from the semiclassical result (Eqn. \ref{scale}). 
Unfortunately, the numerical accuracy for higher levels is not very good, 
which at this stage precludes their interpretation; this is further complicated 
by the smallness of the gaps, which means that the volumes we could reach are 
in fact very small in terms of the correlation length, therefore sizable 
exponential finite size effects are expected. 

For $k=4$, the two-particle gap 
can be seen to be of order $0.9$ from the lowest level above the one-particle 
state; this is roughly consistent with the gap estimate above, but cannot be 
trusted to be of the same precision as there are clear signals of large 
residual cut-off dependence in the behaviour of the level, e.g. the fact 
that it curves upwards for larger values of $mR$, while in reality 
a two-particle level must approach the threshold from above with a 
behaviour $(mR)^{-2}$. For the case $k=6$, the truncation achieved is 
very low and so the higher levels cannot be trusted, precluding their 
analysis for the time being.

\subsection{$\lambda>0$, odd $k$}

\begin{figure}
\begin{centering}
\psfrag{E}{$E/M$}
\psfrag{L}{$MR$}
\includegraphics{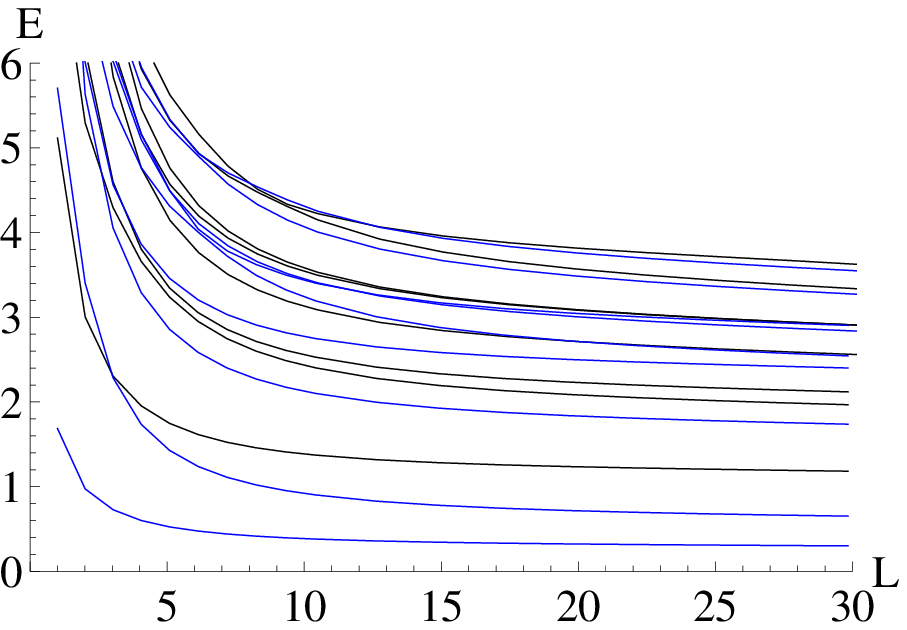} \includegraphics{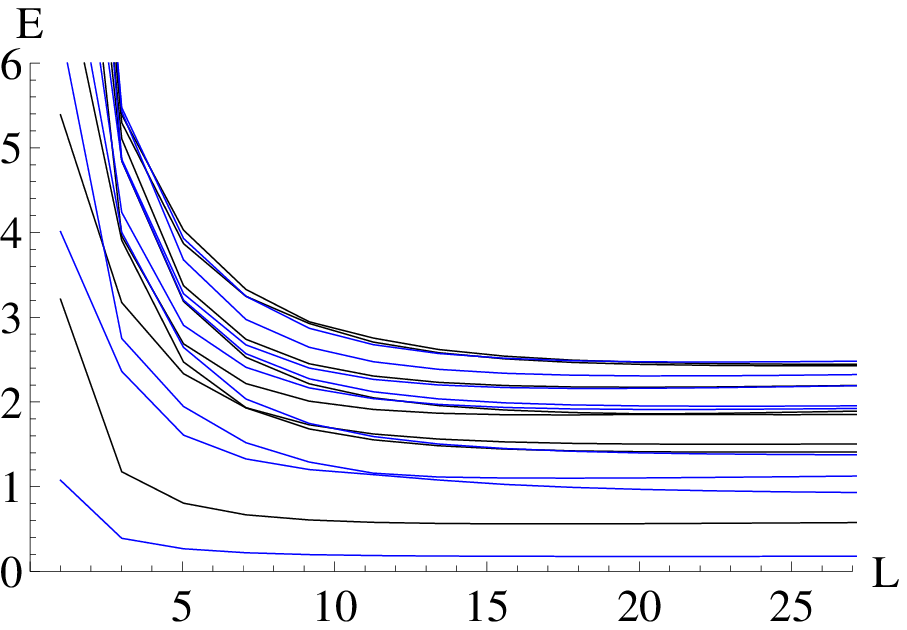}
\par\end{centering}

\protect\caption{\label{fig:gapless}Spectra for $k=3$, $5$ for positive coupling
constant at cut-offs $N=8$ and $6$. Data was improved by the
RC method. Colors represent level data obtained in the integer (black)
and half integer (blue) sectors.}
\end{figure}

In this regime, as we discussed in Section II, we expect a massless flow to an $SU(2)_{1}$ low-energy
fixed point. The mass scale (Eqn. \ref{scale}) in this case corresponds
to the cross-over scale. The spectra shown in Fig. \ref{fig:gapless}
do show marked differences from the even $k$ case. All levels are
monotonically decreasing with the volume, rather than leveling out
as in a massive spectrum. Also, one would expect the gap for $k=3$,
$5$ larger than for $k=4$, $6$, respectively, while compared to
the data in Fig. \ref{fig:TCSAposeven} one readily sees that the
distance between the ground state and the first excited state is,
instead, markedly smaller and monotonically decreasing even at the
largest volume shown.

\begin{figure}
\begin{centering}
\psfrag{E}{$D$}
\psfrag{L}{$MR$}
\includegraphics{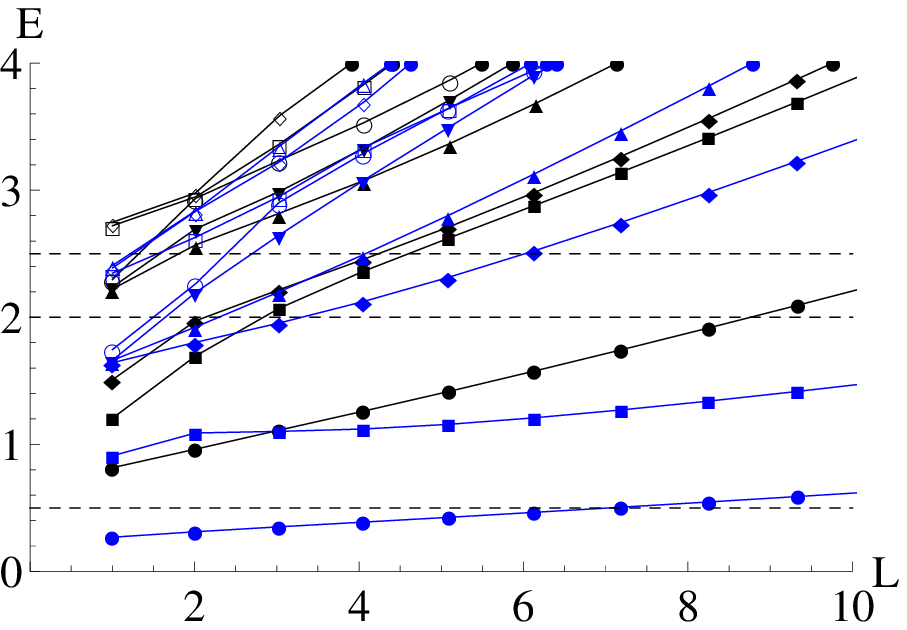} \includegraphics{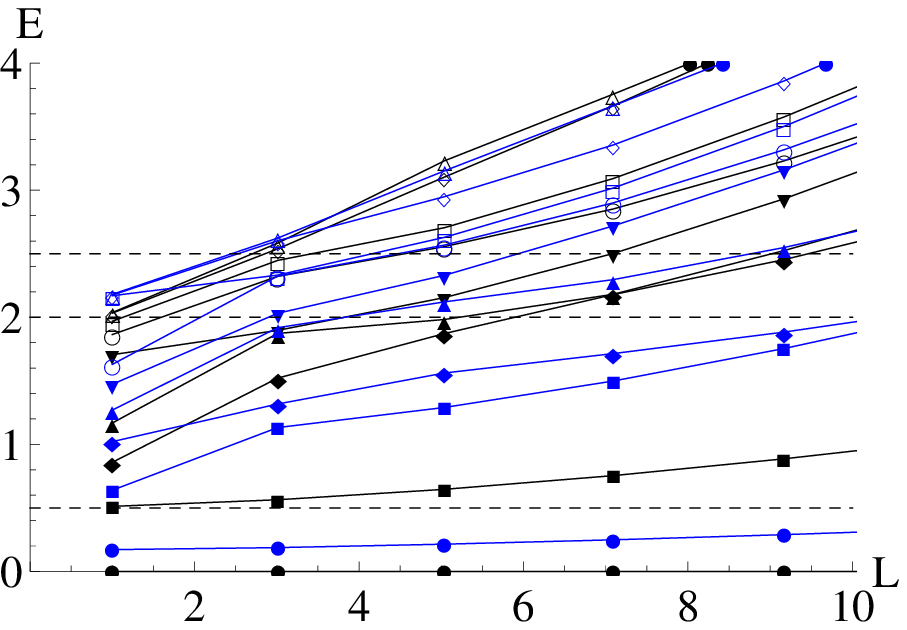}
\par\end{centering}

\protect\caption{\label{fig:scaling_fun}Same data as in Fig. \ref{fig:gapless}, but here we show the
scaling functions, $D_{i}=\frac{R}{2\pi}\left(E_{i}-E_{0}\right)$.
The dashed lines are the first few scaling weights in the $SU(2)_{1}$
CFT.}
\end{figure}

The existence of an infrared fixed point implies that for large values 
of the volume the energy levels should behave as
\[
E_{i}-E_{0}\sim\frac{2\pi x_{i}}{R}+\dots
\]
where $x_{i}$ are scaling dimensions in the $SU(2)_{1}$ theory,
and the dots indicate corrections to the low-energy scaling limit.
One can define the scaling functions
\[
D_{i}=\frac{R}{2\pi}(E_{i}-E_{0})=x_{i}+\dots
\]
As shown in Fig. \ref{fig:scaling_fun}, the detailed matching is
rather limited. There are reasons for which this is expected. First,
from the gaps measured for even $k$, the typical scale parameter for the 
cross-over is expected to be $MR\gtrsim4$. In addition, the 
cross-over itself is slow, due to the
fact that the irrelevant perturbation describing the incoming direction
in the infrared is the current-current perturbation of $SU(2)_{1}$,
which is only marginally irrelevant and leads to a logarithmic approach
to the fixed point \cite{cardy}. Therefore one expects that
the fixed point would only be observable for volume values much higher
than allowed by TCSA accuracy. 

It has long been known that observing an infrared fixed point
in TCSA is difficult \cite{lassig}. In Ref. \cite{lassig} an attempt was made to observe
the flow from the tricritical Ising conformal minimal model to the Ising minimal model by 
perturbing the tricritical Ising theory with the subleading energy perturbation, with 
the conclusion that the behaviour of the first excited state was not inconsistent with 
the existence of the fixed point. 
In a study of two-frequency sine-Gordon model \cite{double2}, the Ising fixed point was 
just barely in the reach of the TCSA. In that case however, the sine-Gordon 
frequency provided a parameter which could be tweaked to improve convergence to the point
that the first two scaling dimensions of the infrared fixed point could be extracted, albeit 
with considerable errors. Note also that in these examples the approach to the fixed point 
was much faster (given by power corrections).

Pending more accurate TCSA numerics (which would require a more accurate
modeling of the cut-off dependence, and more computing power to allow
higher truncation levels), we can only say that the TCSA data are
qualitatively consistent with the existence of a low-energy 
quantum critical point, and the first scaling function in Fig. \ref{fig:scaling_fun}
is also roughly consistent with the lowest scaling weight 
\[
x_{1}=\frac{1}{2}.
\]

\subsection{\label{subsec:k2}The case $k=2$}

\begin{figure}
\begin{centering}
\psfrag{E}{$E/M$}
\psfrag{L}{$MR$}
\includegraphics[scale=1]{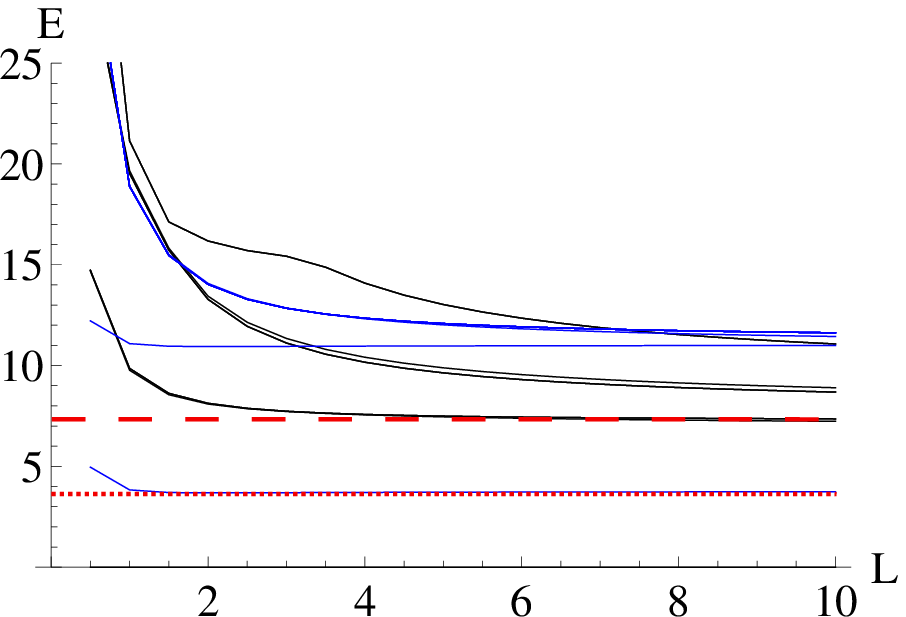}\\
(a) $\lambda>0$\\
\includegraphics[scale=1]{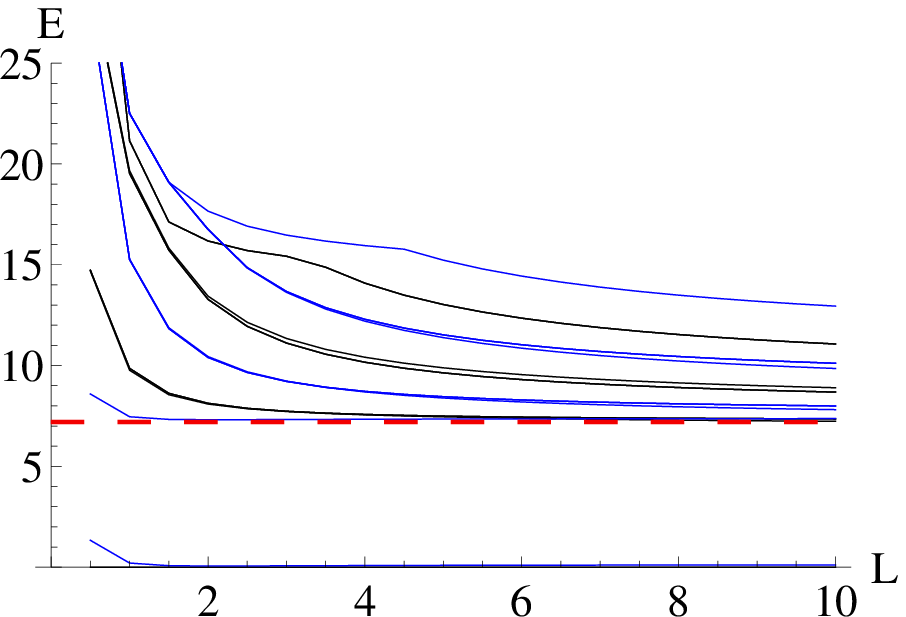}\\ 
(b) $\lambda<0$
\par\end{centering}
\protect\caption{\label{fig:k2}Spectra for $k=2$. The red dotted line (shown only for $\lambda>0$) 
corresponds to the fermion mass $m=2\pi M/\sqrt{3}$, while the red dashed lines show the position of
the two-particle threshold $2m$.}
\end{figure}

For this case one expects a spectrum of free massive Majorana fermions; in our units given by 
Eqn. \ref{eq:mass_scale} 
the fermions mass is
\begin{equation}
 m=\frac{2\pi}{\sqrt{3}}M.
\end{equation}
The resulting spectra for $\lambda>0$ and $\lambda<0$ are shown in Fig. \ref{fig:k2}. They are 
exactly the spectra expected for three Majorana fermions in the $\mathbb{Z}_2$ symmetric and 
$\mathbb{Z}_2$ symmetry breaking phases, respectively. Note that in the $\lambda<0$ the 
excitations are kinks, therefore there are no single-particle levels and every multi-kink 
level appears in two copies (one even, the other odd), which are split by tunneling effects 
decaying exponentially with volume. There are no kink-antikink bound states below the 
two-particle threshold. For the $\lambda<0$, the $k=2$ point is analogous to the 
free-fermion point of sine-Gordon theory, while the $k>2$ cases correspond to 
attractive regime. The main difference from the sine-Gordon case is the non-integrability 
and the presence of $SU(2)$ invariance.

In the symmetric phase $\lambda>0$, there is a unique vacuum and 
the even/odd sectors correspond to levels with even/odd fermion numbers.  In particular, 
the odd sector does contain one-particle states. 

\section{Summary and Conclusions}

We have studied the $SU(2)_k + {\rm Tr}\Phi_{adj}$ theory using the TCSA+RG approach.  We have compared our numerical results to the semiclassical
analysis of this model.
We recall that the semiclassical picture suggests two regimes. One of them corresponds to the negative sign of the coupling and has a doubly degenerate ground state 
characterized by a nonzero vacuum average of the SU(2) matrix field $\la \mbox{Tr}g\ra = \pm \s$. Therefore one expects kinks interpolating between the different vacua. 
Indeed TCSA data show a spectrum that resembles those in the $\Phi^4$ and also the 2-folded sine-Gordon theory, where particles arise as bound states of kinks with a spectrum 
that becomes dense in the semiclassical limit which corresponds here to large $k$. 
The lowest lying excited states consist of two triplets 
of particles with the same mass and can be identified with the lightest kink-antikink bound states. 

In the other regime for even $k$ the semiclassical considerations suggest a triplet of massive excitations, whose low-energy dynamics is  governed by
the $O(3)$ sigma model. The presence of a single vacuum with a massive triplet is confirmed by TCSA, but the precision is significantly smaller since due to the 
much smaller gap, it is necessary to go to much larger volumes which increases the truncation effects. Even so, the vacuum energy density and the mass gap can still 
be extracted with a reasonable accuracy by applying renormalization group improvement techniques. For odd $k$, the semiclassical considerations imply the presence 
of a low-energy quantum critical point described by the $SU(2)_1$ conformal field theory. The TCSA data are consistent with this prediction, but are not accurate enough to 
identify the nature of the fixed point conclusively. We still think that the semiclassical picture has proven robust enough so that this prediction can be trusted.

For the exactly solvable case $k=2$ the spectrum of the model describes three Majorana fermions which also constitute a triplet. 
The model is then non-interacting, therefore there are no more particles in the spectrum. For the two signs of the coupling the spectrum 
differs as usual for a free fermion theory, and fits well into the pattern observed for $k>2$.

\appendix

\section{Realization of $SU(2)_k + Tr \Phi_{adj}$}\label{sec:appendix}

Here we give an example of how the perturbed WZNW model of Eqn. \ref{action} 
can appear in the context of an electronic model.  
We start with a lattice model with U(1)$\times$SU(2)$\times$SU(k) symmetry,
\begin{eqnarray}
 H & = & \sum_n\Big\{-t\Big[\psi^\dagger_{j\s}(n+1)\psi_{j\s}(n) + H.c.\Big] +
U\sum_{\{j\s\} \neq
  \{i\s'\}}[\psi^\dagger_{j\s}(n)\psi_{j\s}(n)][\psi^\dagger_{i\s'}(n)\psi_{i\s'}(n)]
\nonumber\\  
& & - J \psi^\dagger_{j\s}(n)\psi_{i\s}(n)\psi^\dagger_{i\s'}(n)\psi_{j\s'}(n)\Big\}, \label{lattice}
\end{eqnarray}
where $\psi^\dagger_{j\s}(n)$ and $\psi_{j\s}(n)$ are creation and annihilation operators of electrons 
located at sites $n$; $\s = \pm 1/2$ are spin and $i,j = 1,...k$ are orbital indices. 
Treating the interaction as small in comparison with the Fermi energy and assuming that the band 
is far from being half filled, we separate fast and slow Fourier harmonics of the electron operators:
\bea
\psi(n) = \re^{-i k_F na_0}R(x) + \re^{i k_F na_0}L(x), ~~ x = na_0,
\eea
where $k_F$ is the Fermi wave vector and $a_0$ is the lattice constant, and arrive to the continuum version of Eqn. \ref{lattice} in the form of the chiral Gross-Neveu model  
with the most general current-current interaction. The corresponding Hamiltonian density is 
\bea  
&& {\cal H} = \ri v(- R^\dagger_{\s j}\p_xR_{\s j} + L^\dagger_{\s j}\p_x L_{\s j}) + g_cR^\dagger_{\s j}R_{\s j}L^\dagger_{\s' p}L_{\s' p} + \label{Model}\\
&& g_{o}R^\dagger(\tau^a\otimes I)R L^\dagger(\tau^a\otimes I)L + g_{so}R^\dagger(\tau^a\otimes \s^b)R L^\dagger(\tau^a\otimes \s^b)L 
+ g_sR^\dagger(I\otimes \s^a)R L^\dagger(I\otimes \s^a)L, \nonumber
\eea
where $\s^a$ ($a=1,2,3$) acting on the Greek indices and $\tau^a$ ($a=1,...k^2-1$) acting on the Latin indices are generators of the su(2) and su(k) algebras respectively, normalized as 
\[
Tr(\s^a\s^b) = Tr(\tau^a\tau^b) = \frac{\delta_{ab}}{2}.
\]
The coupling constants $g_{1,2,3}$ are related to $U$ and $J$ while $v$ the Fermi velocity is given by $v = 2t\sin(k_Fa_0)$.
 
Model (Eqn. \ref{Model}) is integrable for $g_o=g_{so}/2,  g_s= g_{so}/k$ where in the case the symmetry expands
to U(1)$\times$SU(2k).  In this case the abelian sector is massless and the 
non-abelian sector is massive if at least one of $g_{o}$ or $g_{s}$ is positive and $g_{so} \neq 0$ and is massless 
otherwise. It is also integrable if $g_{so} =0$. For this last case the Hamiltonian density can be written as a sum of 
three independent Wess-Zumino-Novikov-Witten (WZNW) models perturbed by the current-current interactions:
\bea
&& {\cal H} = \Big[\frac{2\pi}{k+2}\Big(:J_R^aJ_R^a: +:J_L^aJ_L^a: \Big) + g_sJ_R^aJ_L^a\Big] + \\
&& \Big[\frac{2\pi}{k+2}\Big(:F_R^aF_R^a: +:F_L^aF_L^a: \Big) + g_oF_R^aF_L^a\Big] + \\
&& \Big[\frac{\pi}{k}\Big(:j_Rj_R: +:j_Lj_L: \Big) + g_cj_Rj_L\Big],
\eea
where $J^a_{R/L}$, $F^a_{R/L}$, and $j_{R/L}$  are $SU(2)_k$, $SU(k)_2$ and $U(1)$ left/right Kac-Moody currents.
Each perturbed WZNW model is exactly solvable \cite{tsv87,smirnov}. 

We consider the case $g_s <0$ and small $g_{so}$ when the term mixing the spin and the orbital sectors
in Eqn. \ref{Model} can be considered as a perturbation around the $SU(2)_k$ WZNW critical point. As we shall demonstrate, this perturbation 
is always relevant and is given by the $SU(2)_k$ adjoint operator. 

The standard analytic approach to the models of type (Eqn. \ref{Model}) starts with RG equations. 
On this basis certain robust predictions have been made \cite{fisher,kon_rgflow}.  In particular it has been argued that at the
lowest energies the largest possible symmetry is restored (in our case it would be U(1)$\times$SU(2k)).  We do note that the 
reliability of such approach hinges
in part on weak coupling as the Gell-Mann-Low function is universal only at first loop.  The first loop RG equations 
for the model in Eqn. \ref{Model} are \cite{kotliar2} 
\bea
\dot g_o = kg_o^2 + 3kg_{so}^2/4, ~~ \dot g_{so} = (k-2)g_{so}^2 + g_{so}(kg_s + 2g_o), ~~ \dot g_s = 2g_s^2 +2(k-1/k)g_{so}^2. \label{RG1}
\eea
The case we are interested in is $g_o(0) >0, g_s(0)<0$. Then at $g_{so} =0$ the current-current interaction in the SU(2) invariant 
sector scales to zero and this sector is gapless.  On the other hand, the interaction in the SU(k) sector (the orbital one) scales to strong coupling 
and the excitations in this sector become massive. This occurs at the RG scale $\xi_o \approx 1/g_o(0)k$. At finite $g_{so}$ the 
corresponding term acts as a relevant perturbation. Assuming that $g_{so}$ remains the smallest in flowing to the scale governed by $\xi_o$, 
we extract from Eqn. \ref{RG1} its value at this scale to be:
\be
 g_{so}(\xi_o) \approx \frac{g_{so}(0)}{g_o(0) + 2|g_s(0)|/k}.
\ee
We will assume that $|g_{so}(\xi_o)| \ll 1$ and consider the spin-orbit current-current interaction as a perturbation. 
As it was demonstrated in \cite{akhanjee}, this perturbing operator is the trace of the primary field $\Phi^{adj}_{AB}$ 
of the $SU(2)_k$ WZNW model where  $\Phi^{adj}_{AB}$ is the field belonging to the adjoint representation. This argument is based on 
the observation that the marginally relevant term with $g_{os}$ having scaling dimension 2 can be represented as a product of 
conformal blocks of the $SU(k)_2$ and $SU(2)_k$ primary fields in the adjoint representation. This suggestion is based on their scaling dimensions: 
\bea
d_{adj}[SU(k)_2] = \frac{2k}{k+2}, ~~d_{adj}[SU(2)_k] = \frac{4}{k+2},
\eea
In the vacuum of the perturbed $SU(k)_2$ WZNW theory, only the Tr$\Phi^{adj}$ has a nonzero average. 
This leads one to the conclusion that after the high energy degrees of freedom of this theory are integrated
out, the local operator Tr$\Phi^{adj}[SU(2)_k]$ will emerge from the product of the corresponding conformal blocks. 

We thus estimate the coupling of the perturbation to be equal to
\be
\lambda \sim g_{so}(\xi_0)\la\mbox{Tr}\Phi^{adj}[SU(k)]\ra. \label{lambda}
\ee
As is shown in the main text, the physics of the model in Eqn. \ref{action} depends crucially on the sign of $\lambda$. 
If we consider the WZNW action in Eqn. \ref{action} as a descendant of the fermionic model (Eqn. \ref{Model}), 
the sign is determined by the product of signs of $g_{so}$ and the vacuum average of the adjoint operator in the SU(k) 
sector (Eqn. \ref{lambda}). The ground state of the $SU(k)_2$ model perturbed by the current-current interaction 
is degenerate and hence the magnitude and the sign of $\lambda$ depend on the vacuum. 
This degeneracy is lifted by the interaction with the $SU(2)$ sector.  As a result the sign of the 
interaction (Eqn. \ref{lambda}) must be chosen so as to minimize the ground state energy. 
As we have demonstrated (cf. Table \ref{tab:bulk}), for a given $k$ the ground state energy is always lower for $\lambda <0$ 
where the masses in the $SU(2)$ sector are the largest and the lowest energy excitations consist of a massive triplet of particles, 
which appear in two copies due to the degenerate pair of vacua.

\subsubsection*{Acknowledgments}
RMK and AMT are supported by the US DOE under contract number DE-AC02-98 CH 10886.  RMK also
received support from by the National Science Foundation under grant no. PHY 1208521.  GT and PT
are supported  by  Hungarian Academy of Sciences both by Momentum grant LP2012-50/2013 and a postdoctoral fellowship for PT.

\end{document}